\documentclass{ws-ijmpa}
\usepackage{url}

\usepackage{tabularx}
\newcommand{\si}{\sigma}

\newcommand{\NLO}{\mathrm{NLO}}

\newcommand{\TeV}{\unskip\,\mathrm{TeV}}

\def\nn{\nonumber}

\def\text{\textstyle}

\def\bc{\begin{center}}
\def\ec{\end{center}}
\def\bi{\begin{itemize}}
\def\ei{\end{itemize}}

\newcommand{\ord}{{\cal O}}
\begin{document}

\markboth{Wim Beenakker {\it et al.}}
{Squark and gluino hadroproduction}

%
%

\title{SQUARK AND GLUINO HADROPRODUCTION}

\author{WIM BEENAKKER}

\address{Theoretical High Energy Physics, IMAPP, Radboud University Nijmegen, P.O. Box 9010\\
  NL-6500 GL Nijmegen, The Netherlands\\
  W.Beenakker@science.ru.nl}

\author{SILJA BRENSING}

\address{DESY, Theory Group
  Notkestrasse 85, D-22603 Hamburg, Germany\\
  silja.christine.brensing@desy.de}

\author{MICHAEL KR\"AMER}

\address{Institute for Theoretical Particle Physics and Cosmology, RWTH Aachen University\\
  D-52056 Aachen, Germany\\
  mkraemer@physik.rwth-aachen.de}

\author{ANNA KULESZA}

\address{Institute for Theoretical Particle Physics and Cosmology, RWTH Aachen University\\
  D-52056 Aachen, Germany\\
  anna.kulesza@physik.rwth-aachen.de}

\author{ERIC LAENEN}

\address{ITFA, University of Amsterdam, Science Park 904, 1090 GL Amsterdam, \\
  ITF, Utrecht University, Leuvenlaan 4, 3584 CE Utrecht\\
  Nikhef, Science Park 105, 1098 XG Amsterdam, The
  Netherlands\\
  Eric.Laenen@nikhef.nl}

\author{LESZEK MOTYKA}

\address{ Institute of Physics,
  Jagellonian University, Reymonta 4, 30-059
  Krak\'ow, Poland\\
  leszekm@th.if.uj.edu.pl}

\author{IRENE NIESSEN}

\address{Theoretical High Energy Physics, IMAPP, Radboud University Nijmegen, P.O. Box 9010\\
  NL-6500 GL Nijmegen, The Netherlands\\
  i.niessen@science.ru.nl}

\maketitle


\begin{abstract}
  We review the theoretical status of squark and gluino
  hadroproduction and provide numerical predictions for all squark and
  gluino pair-production processes at the Tevatron and at the LHC,
  with a particular emphasis on proton-proton collisions at 7 TeV. Our
  predictions include next-to-leading order supersymmetric QCD
  corrections and the resummation of soft gluon emission at
  next-to-leading-logarithmic accuracy. We discuss the impact of the
  higher-order corrections on total cross sections, and provide an
  estimate of the theoretical uncertainty due to scale variation and
  the parton distribution functions.  
\end{abstract}

\ccode{PACS numbers: 11.25.Hf, 123.1K}

\ccode{Preprint numbers: DESY 11-060, ITFA 11-09, ITP-UU-11/13, Nikhef-2011-011, TTK 11-12}

\section{Introduction}	

The search for supersymmetry (SUSY)\cite{Golfand:1971iw,Wess:1974tw}
is a central part of the experimental program at the hadron colliders
Tevatron and LHC. Models of weak-scale SUSY provide a promising
solution to the hierarchy problem of the Standard Model (SM) and
comprise new supersymmetric particles (sparticles) with masses of
order 1 TeV. The coloured TeV-scale sparticles, squarks ($\tilde q$)
and gluinos ($\tilde g$), would be produced copiously in hadronic
collisions and thus offer the strongest sensitivity for supersymmetry
searches at the Tevatron and the LHC.

We consider the minimal supersymmetric extension of the Standard Model
(MSSM)\cite{Nilles:1983ge,Haber:1984rc} where, as a consequence of
R-parity conservation, squarks and gluinos are pair-produced in
collisions of two hadrons $h_1$ and $h_2$:
\begin{equation}
  h_1 h_2 \;\to\; \tilde{q}\tilde{q}\,,
  \tilde{q}\bar{\tilde{q}}\,, \tilde{q}\tilde{g}\,, \tilde{g}\tilde{g},\, 
  \tilde t_1\bar{\tilde t}_1,\, \tilde t_2\bar{\tilde t}_2 + X\,.
\label{eq:processes}
\end{equation}
The production of top squarks (stops), $ \tilde t_{1,2}$, has to be
treated separately, because the strong Yukawa coupling between top
quarks, stops and Higgs fields gives rise to potentially large mixing
effects and mass splitting.\cite{Ellis:1983ed} In
Eq.~(\ref{eq:processes}) and throughout the rest of this paper,
$\tilde t_1$ and $\tilde t_2$ denote the lighter and heavier stop mass
eigenstate, respectively. For the other squarks we suppress the
chiralities, i.e.\ $\tilde{q} =(\tilde{q}_{L}, \tilde{q}_{R})$, and do
not explicitly state the charge-conjugated processes.

Searches for squarks and gluinos at the proton--antiproton collider
Tevatron with a centre-of-mass energy of $\sqrt{S}=1.96$~TeV have
placed lower limits on squark and gluino masses in the range of 300 to
400\,GeV, depending in detail on the specific SUSY
model.\cite{:2007ww,Aaltonen:2008rv} The proton--proton collider LHC,
which has been operating at $\sqrt{S}=7$~TeV in 2010, has already
significantly extended the squark and gluino mass limits to values of
around 850\,GeV.\cite{Khachatryan:2011tk}%
\cdash\cite{Aad:2011ks} Dedicated searches for the lighter stop mass
eigenstate at LEP\cite{LEP-SUSY,Heister:2002hp} and the
Tevatron\cite{Aaltonen:2007sw}\cdash\cite{Abazov:2008kz} have placed
lower limits in the range 70 to 200\,GeV. Already in 2011, with a
projected integrated luminosity of 1 to 2\,fb$^{-1}$, the LHC should
be sensitive to squarks and gluinos with masses in the TeV
region\cite{Bechtle:2011dm}, while SUSY particles with masses up to
3\,TeV can be probed once the LHC reaches its design energy of
$\sqrt{S}=14$~TeV.\cite{Aad:2009wy,Bayatian:2006zz}

Accurate theoretical predictions for inclusive squark and gluino cross
sections are needed both to set exclusion limits and, in case SUSY is
discovered, to determine SUSY particle masses and
properties.\cite{Baer:2007ya}\cdash\cite{Dreiner:2010gv} The inclusion
of higher-order SUSY-QCD corrections significantly reduces the
renormalization- and factorization-scale dependence of the
predictions.  In general, the corrections also increase the size of
the cross section with respect to the leading-order
prediction\cite{Kane:1982hw}\cdash\cite{Dawson:1983fw} if the
renormalization and factorization scales are chosen close to the
average mass of the produced SUSY particles. Consequently, the
SUSY-QCD corrections have a substantial impact on the determination of
mass exclusion limits and would lead to a significant reduction of
errors on SUSY mass or parameter values in the case of discovery.  The
processes listed in Eq.~(\ref{eq:processes}) have been known for quite
some time at next-to-leading order (NLO) in
SUSY-QCD.\cite{Beenakker:1994an}\cdash\cite{Beenakker:1997ut}
Electroweak corrections to the $\ord (\alpha_{\rm s}^2)$ tree-level
processes\cite{Hollik:2007wf}\cdash\cite{Germer:2011an} and the
electroweak Born production channels of $\ord (\alpha\alpha_{\rm s})$
and $\ord (\alpha^2)$\cite{Alan:2007rp,Bornhauser:2007bf} are in
general significant for the pair production of SU(2)-doublet squarks
$\tilde{q}_L$ and at large invariant masses, but they are moderate for
inclusive cross sections.

A significant part of the NLO QCD corrections can be attributed to the
threshold region, where the partonic centre-of-mass energy is close to
the kinematic production threshold. In this region the NLO corrections
are dominated by soft gluon emission off the coloured particles in the
initial and final state and by the Coulomb corrections due to the
exchange of gluons between the massive sparticles in the final state.
The soft-gluon corrections can be taken into account to all orders in
perturbation theory by means of threshold resummation
techniques.\cite{Sterman:1986aj,Catani:1989ne}

Recently, such a threshold resummation has been performed for all MSSM
squark and gluino production processes, Eq.\,(\ref{eq:processes}), at
next-to-leading-logarithmic (NLL)
accuracy.\cite{Kulesza:2008jb}\cdash\cite{Beenakker:2010nq} A
formalism has been developed in the framework of effective field
theories which allows for the resummation of soft and Coulomb gluons
in the production of coloured sparticles, but has so far only been
applied to squark-antisquark
production.\cite{Beneke:2009rj,Beneke:2010da} In addition, the
dominant next-to-next-to-leading order (NNLO) corrections, including
those coming from the resummed cross section at
next-to-next-to-leading-logarithmic (NNLL) level, have been calculated
for squark-antisquark pair-production.%
\cite{Langenfeld:2009eg,Langenfeld:2010vu} The production of gluino
bound states as well as bound-state effects in gluino-pair production
has also been studied.\cite{Hagiwara:2009hq,Kauth:2009ud}

In this work we will present the state-of-the-art SUSY-QCD predictions
for the MSSM squark and gluino hadroproduction processes,
Eq.~(\ref{eq:processes}), at the Tevatron and the LHC, including NLO
corrections and NLL threshold resummation.  The processes will all be
treated on the same footing, i.e.\ classes of beyond-NLO effects that
have been calculated only for one particular process or reaction
channel will not be taken into account. We will discuss the impact of
the SUSY-QCD corrections on the total cross sections and provide an
estimate of the theoretical uncertainty due to scale variation, parton
distribution functions, and the strong coupling $\alpha_{\rm s}$.

The structure of the paper is as follows. In
section~\ref{sec:nll_resummation} we briefly review the application of
the resummation technique to total cross sections for coloured
sparticle pair-production. The numerical results are presented in
section~\ref{sec:numerics}. We show predictions for the Tevatron, and
for the LHC with centre-of-mass energies of $\sqrt{S}=7$~TeV and
$\sqrt{S}=14$~TeV. We will conclude in section~\ref{sec:conclusions}.

\section{NLL resummation}
\label{sec:nll_resummation}

In this section we provide a brief background to the calculation of
the threshold-resummed cross
sections\cite{Kulesza:2008jb}\cdash\cite{Beenakker:2009ha} we use for
our results in the next section.  The resummation for $2\to 2$
processes with all four external legs carrying colour has been studied
extensively in the literature, specifically for
heavy-quark\cite{Kidonakis:1997gm,Bonciani:1998vc} and jet
production.\cite{Kidonakis:1998bk}\cdash\cite{Bonciani:2003nt} In our
calculations we make use of the framework developed there.

The hadronic threshold for the inclusive production of two final-state
particles $k, l$ with masses $m_k$ and $m_l$ corresponds to a hadronic
centre-of-mass energy squared that is equal to $S=(m_k+m_l)^2$.  Thus
we define the threshold variable $\rho$, measuring the distance from
threshold in terms of energy fraction, as
\begin{equation}
  \label{eq:1}
\rho \;=\; \frac{(m_k+m_l)^2}{S}\,.  
\end{equation}
Our results are based on the following expression for the NLL-resummed
cross section, matched to the exact NLO calculation%
\cite{Beenakker:1994an}\cdash\cite{Beenakker:1997ut}
\begin{eqnarray}
\label{eq:14}
\si^{\rm (NLO+NLL)}_{h_1 h_2 \to kl}\bigl(\rho, \{m^2\},\mu^2\bigr) 
  &=& \si^{\rm (NLO)}_{h_1 h_2 \to kl}\bigl(\rho, \{m^2\},\mu^2\bigr)\nn
          \\[1mm]
   &&  \hspace*{-30mm}+\, \frac{1}{2 \pi i} \sum_{i,j=q,\bar{q},g}\, \int_\mathrm{CT}\,dN\,\rho^{-N}\,
       \tilde f_{i/h_1}(N+1,\mu^2)\,\tilde f_{j/h_{2}}(N+1,\mu^2) \nn\\[0mm]
   && \hspace*{-20mm} \times\,
       \left[\tilde\si^{\rm(res)}_{ij\to kl}\bigl(N,\{m^2\},\mu^2\bigr)
             \,-\, \tilde\si^{\rm(res)}_{ij\to kl}\bigl(N,\{m^2\},\mu^2\bigr)
       {\left.\right|}_{\scriptscriptstyle({\NLO})}\, \right]\,,
\end{eqnarray}
where the last term in the square brackets denotes the NLL resummed
expression expanded to NLO. The initial state hadrons are denoted
generically as $h_1$ and $h_2$, and $\mu$ is the common factorization and
renormalization scale. The resummation is performed after
taking a Mellin transform (indicated by a tilde) of the cross section,
\begin{equation}
  \label{eq:10}
  \tilde\si_{h_1 h_2 \to kl}\bigl(N, \{m^2\}\bigr) 
 \equiv \int_0^1 d\rho\;\rho^{N-1}\;
           \si_{h_1 h_2\to kl}\bigl(\rho,\{ m^2\}\bigr) \,.
\end{equation}
To evaluate the contour CT of the inverse Mellin transform in
Eq.~(\ref{eq:14}) we adopt the so-called ``minimal
prescription''.\cite{Catani:1996yz}  The NLL resummed cross section in
Eq.~(\ref{eq:14}) reads
\begin{multline}
  \label{eq:12}
  \tilde{\sigma}^{\rm (res)} _{ij\rightarrow kl}\bigl(N,\{m^2\},\mu^2\bigr) 
= \sum_{I}\,
      \tilde\sigma^{(0)}_{ij\rightarrow kl,I}\bigl(N,\{m^2\},\mu^2\bigr)\, 
      C_{ij \rightarrow kl, I}\bigl(N,\{m^2\},\mu^2\bigr) \\[1mm]
   \times\,\Delta_i (N+1,Q^2,\mu^2)\,\Delta_j (N+1,Q^2,\mu^2)\,
     \Delta^{\rm (s)}_{ij\rightarrow
       kl,I}\bigl(N+1,Q^2,\mu^2\bigr)\,,
\end{multline}
where $\tilde{\sigma}^{(0)}_{ij \rightarrow kl, I}$ are the
colour-decomposed leading-order cross sections in Mellin-moment space,
with $I$ labelling the possible colour
structures.\cite{Kulesza:2009kq,Beenakker:2009ha} 
Here we have introduced the hard scale $Q^2 = (m_k + m_l)^2$. 
The perturbative
functions $C_{ij \rightarrow kl, I}$ contain information about hard
contributions beyond leading order.  This information is only relevant
beyond NLL accuracy and therefore we keep $C_{ij \rightarrow kl,I} =1
$ in our calculations. The functions $\Delta_{i}$ and $\Delta_{j}$ sum
the effects of the (soft-)collinear radiation from the incoming
partons. They are process-independent and do not depend on the colour
structures.  They contain the leading logarithmic dependence, as well
as part of the subleading logarithmic behaviour. The expressions for
$\Delta_{i}$ and $\Delta_{j}$ can be found in the
literature.\cite{Kulesza:2009kq} The resummation of the soft-gluon
contributions, which does depend on the colour structures in which the
final state SUSY particle pairs can be produced, contributes at the
NLL level and is summarized by the factor
\begin{equation}
  \Delta_{I}^{\rm (s)}\bigl(N,Q^2,\mu^2\bigr) 
  \;=\; \exp\left[\int_{\mu}^{Q/N}\frac{dq}{q}\,\frac{\alpha_{\rm s}(q)}{\pi}
                 \,D_{I} \,\right]\,.
\label{eq:2}
\end{equation}
The one-loop coefficients $D_{I}$ follow from the threshold limit of
the one-loop soft anomalous-dimension
matrix.\cite{Kulesza:2009kq,Beenakker:2009ha}

Two remarks can be made regarding squark-gluino production, and
stop-antistop production in particular. The former reaction is the
only one in our set where heavy, coloured SUSY particles of different
mass are produced. Our resummed expressions are sensitive to these
differences through the Born cross sections, and through the resummed
exponents at the NLL level.\cite{Beenakker:2009ha} In stop-antistop
production through $q\bar{q}$ annihilation the gluino exchange
diagram, which would require top parton distribution functions, is
missing in the five-flavour scheme we adopt. As a consequence, the
Born cross section for stop production is proportional to
$\left(\sqrt{1-4m_{\tilde{q}}^2/s}\,\right)^3$, as opposed to
$\sqrt{1-4m_{\tilde{q}}^2/s}$ for production of the other squark
flavours. Here, $m_{\tilde{q}}$ is a generic squark mass and $s$
denotes the partonic centre-of-mass energy squared. We have
argued\cite{Beenakker:2010nq} that this has no effect on the resummed
expression other than through the explicit expression for the Born
function.

\section{Numerical Results}
\label{sec:numerics}

We present numerical predictions for squark and gluino production at
the Tevatron ($\sqrt{S}=1.96$~TeV) and the LHC ($\sqrt{S}=7$ and
14~TeV).  We compare LO, NLO and NLO+NLL matched results, and discuss
the theoretical uncertainty due to the choice of renormalization and
factorization scales and due to the uncertainty in the parton
distribution functions (pdfs) and the QCD coupling $\alpha_{\rm s}$.
The LO and NLO cross sections%
\cite{Beenakker:1994an}\cdash\cite{Beenakker:1997ut} are available in
the form of the public computer code {\tt Prospino}.\cite{prospino}
The $\overline{\rm MS}$-scheme with five active flavours is used to
define $\alpha_{\rm s}$ and the parton distribution functions at
NLO. The masses of the squarks and gluinos are renormalized in the
on-shell scheme, and the SUSY particles are decoupled from the running
of $\alpha_{\rm s}$ and the pdfs.

As mentioned in the introduction, the production of stops has to be
treated separately because of potentially large mixing effects and
mass splitting. The production of the other squark flavours, which we
assume to be mass degenerate, is treated together, i.e. we sum over
five flavours of squarks, $\tilde{q} \in \{\tilde{u}, \tilde{d},
\tilde{c}, \tilde{s}, \tilde{b}\}$. In that case our numerical
predictions include both chiralities ($\tilde{q}_L$ and $\tilde{q}_R$)
and the charge-conjugated processes.

The stop cross section is shown separately. Note that since mixing in
the stop sector enters explicitly only through higher-order diagrams,
the stop-mixing angle $\theta_{\tilde{t}}$ need not be renormalized
and one can use the lowest-order expression derived from the stop mass
matrix. Beyond LO the stop cross section does not only depend on the
stop mass, but also on the gluino mass $m_{\tilde{g}}$, the average
mass of the first and second generation squarks $m_{\tilde{q}}$ and
the mixing angle $\theta_{\tilde{t}}$. We have fixed $m_{\tilde{g}}$,
$m_{\tilde{q}}$ and $\theta_{\tilde{t}}$ according to the SPS1a'
benchmark scenario.\cite{AguilarSaavedra:2005pw} Note, however, that
the dependence of the stop cross section on the SUSY parameters that
enter only at NLO is numerically very small with variations of at most
2\%.\cite{Beenakker:2010nq} The numerical results presented for stop
production also apply to sbottom production when the same input
parameters are adopted, since the impact of bottom-quark induced
contributions to sbottom hadroproduction is
negligible.\cite{Beenakker:2010nq}

For convenience we define the average mass of the final-state
sparticle pair $m = (m_k + m_l)/2$, which reduces to the squark and
gluino mass for $\tilde{q}\bar{\tilde{q}} $, $\tilde{q}\tilde{q}$, and
$\tilde{g}\tilde{g}$ final states, respectively. The renormalization
and factorization scales are taken to be equal, $\mu_{\rm R} =
\mu_{\rm F} = \mu$.  As our default, hadronic NLO and NLO+NLL cross
sections are obtained with the 2008 NLO MSTW pdfs\cite{MSTW:private}
and the corresponding $\alpha_{\rm s}(M_{Z}) = 0.120$. In particular,
all plots show results for the MSTW pdf set.\cite{MSTW:private} We
also present results based on the CT10\cite{Lai:2010vv} and
CTEQ6L1\cite{CTEQ6LO} pdf sets in tables, for comparison.

Let us now discuss the numerical results for squark and gluino
hadroproduction at the Tevatron, and at the LHC operating with 7 and
14\,TeV hadronic centre-of-mass energy. We shall study the scale
dependence of the LO, NLO and NLO+NLL cross sections, the impact of
the NLL threshold resummation, and present our best predictions at
NLO+NLL for the inclusive cross sections, including the theoretical
uncertainties from scale variation as well as the pdf and $\alpha_{\rm
  s}$ errors. We put special emphasis on the predictions for the LHC
at 7\,TeV energy, which are of immediate importance for the upcoming
SUSY searches, and collect detailed results for representative LO, NLO
and NLO+NLL cross sections and the corresponding theory uncertainties
in tables.

\subsection{Tevatron}
\begin{figure}
\begin{center}
\epsfig{file=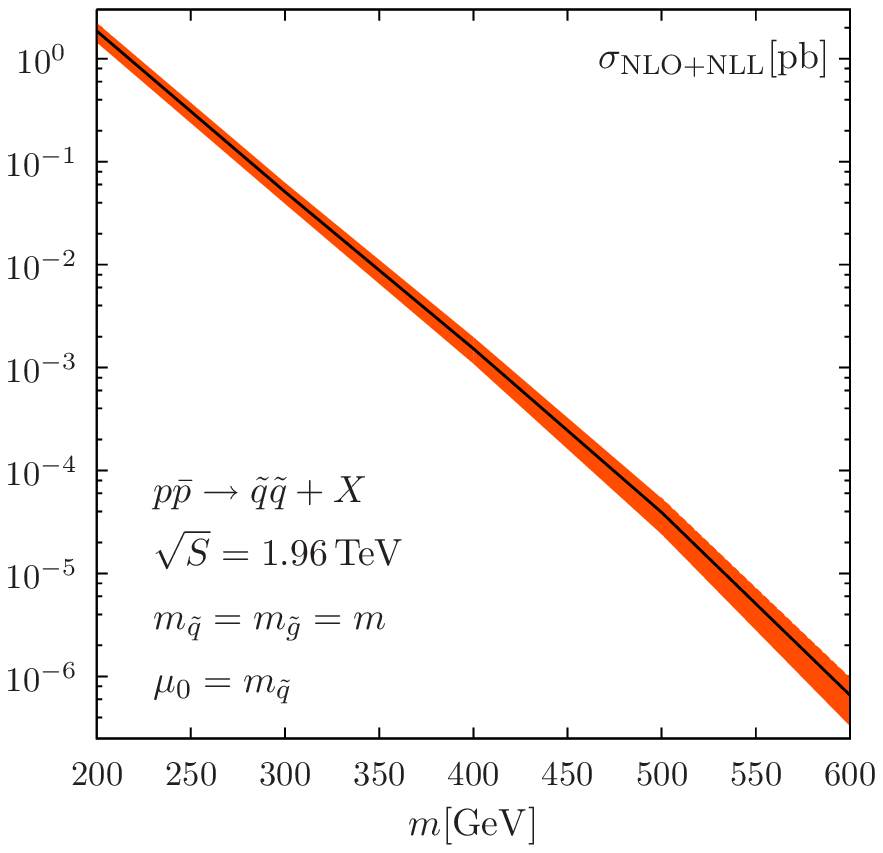,
  width=0.45\columnwidth,angle=0}
\epsfig{file=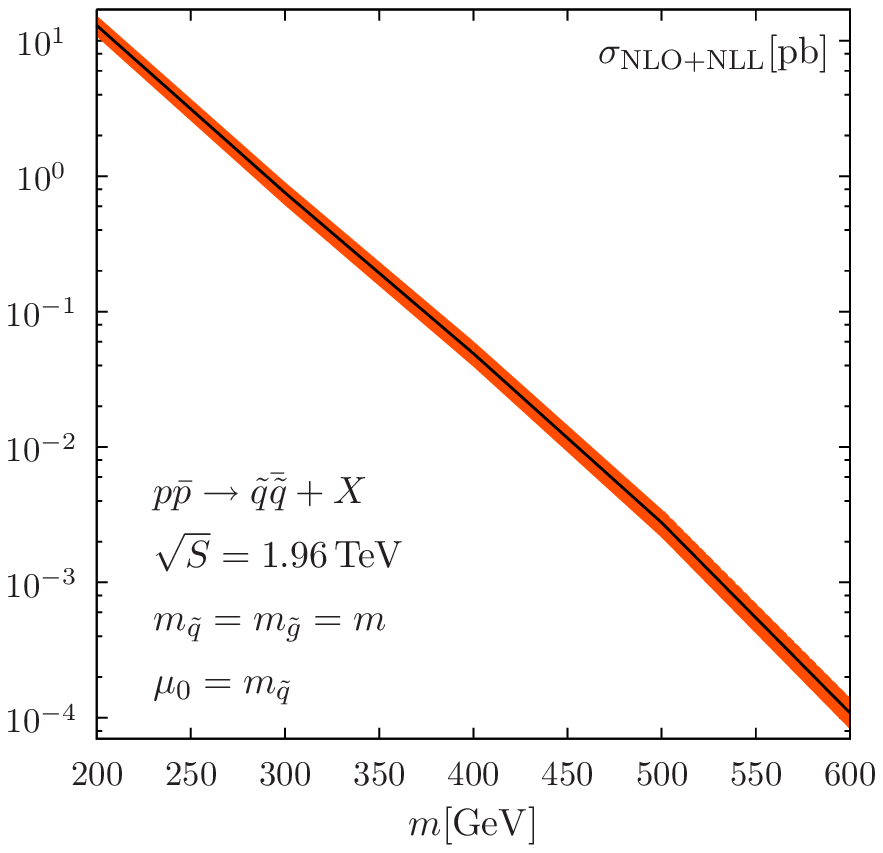,
  width=0.45\columnwidth,angle=0}\\[5mm]
\epsfig{file=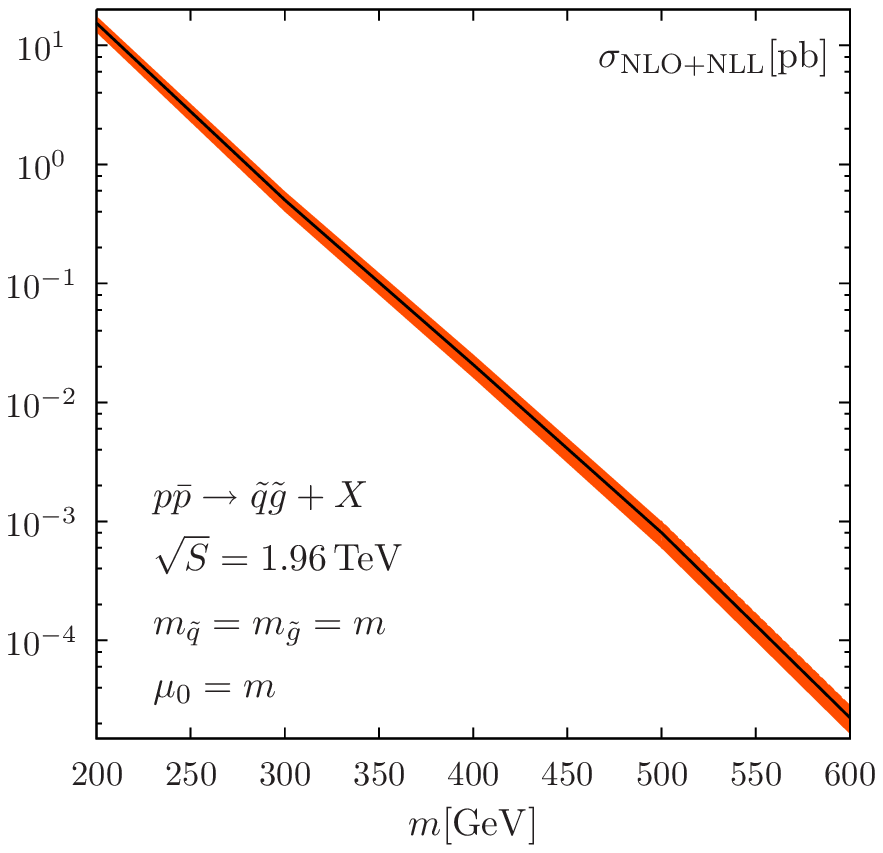,
  width=0.45\columnwidth,angle=0}
\epsfig{file=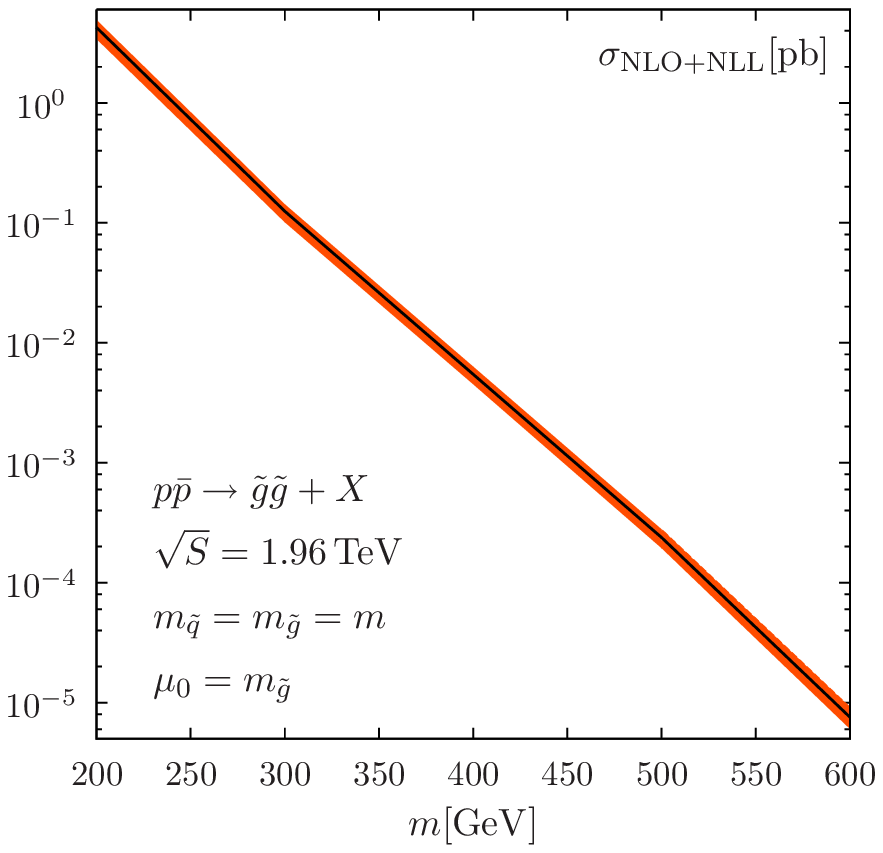,
  width=0.45\columnwidth,angle=0}\\[5mm]
\epsfig{file=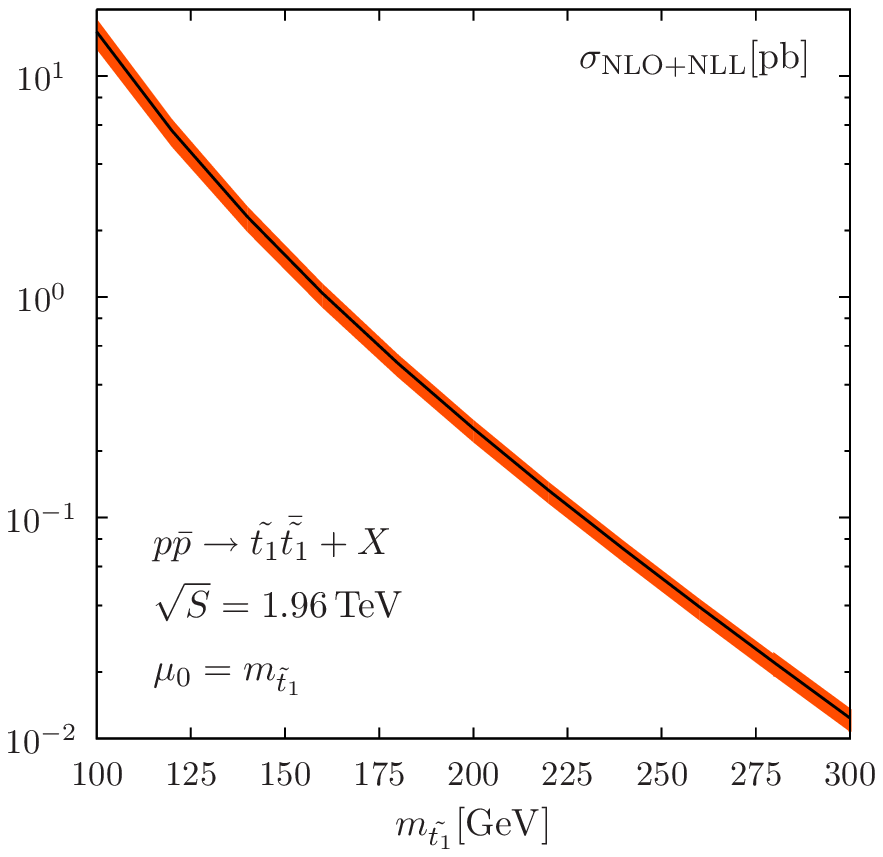, 
  width=0.45\columnwidth,angle=0}
\end{center}
\caption{The NLO+NLL SUSY-QCD cross section for the individual squark
  and gluino pair-production processes at the Tevatron, $p\bar{p}\to
  \tilde{q}\tilde{q}\,, \tilde{q}\bar{\tilde{q}}\,,
  \tilde{q}\tilde{g}\,, \tilde{g}\tilde{g}+X$ and $p\bar{p}\to\tilde
  t_1\bar{\tilde t}_1+ X$, as a function of the average sparticle mass
  $m$. The error band includes the 68\% C.L.\ pdf and $\alpha_{\rm s}$
  error, added in quadrature, and the error from scale variation in
  the range $m/2\le \mu \le 2m$ added linearly to the combined pdf and
  $\alpha_{\rm s}$ uncertainty.\label{fig:xs_tev}}
\end{figure}

\begin{figure}
\begin{center}
\epsfig{file=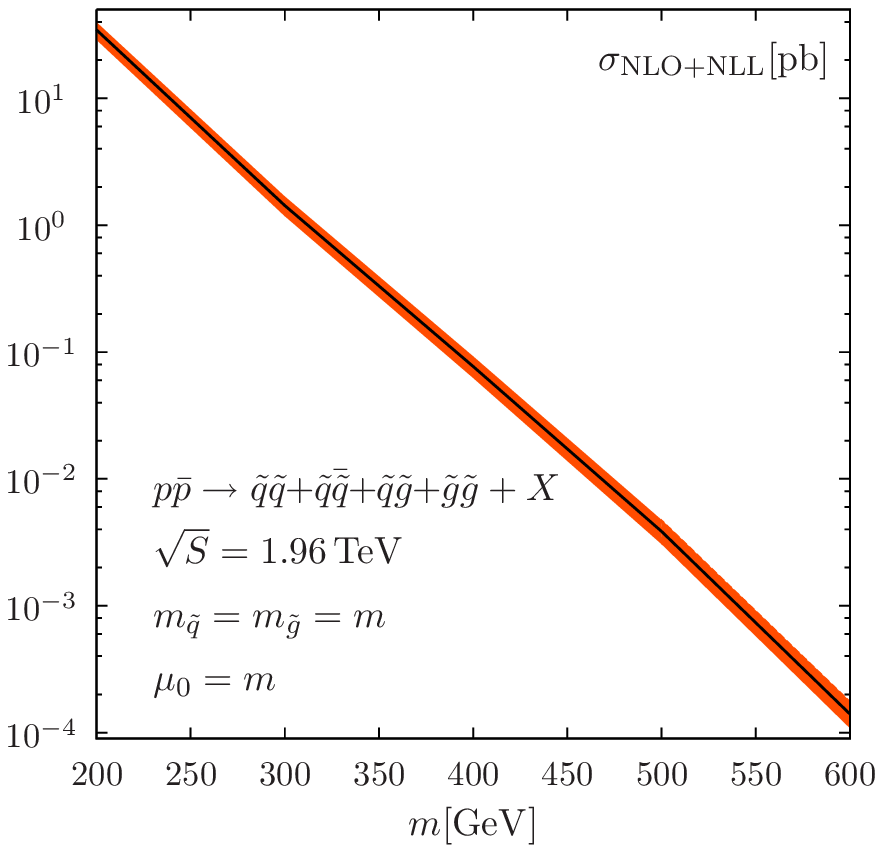,
  width=0.55\columnwidth,angle=0}
\caption{The NLO+NLL SUSY-QCD cross section for inclusive squark and
  gluino pair-production at the Tevatron, $p\bar{p}\to
  \tilde{q}\tilde{q} + \tilde{q}\bar{\tilde{q}} + \tilde{q}\tilde{g}+
  \tilde{g}\tilde{g}+X$, as a function of the average sparticle mass
  $m$. The error band includes the 68\% C.L.\ pdf and $\alpha_{\rm s}$
  error, added in quadrature, and the error from scale variation in
  the range $m/2\le \mu \le 2m$ added linearly to the combined pdf and
  $\alpha_{\rm s}$ uncertainty.\label{fig:xs_sum_tev}}
\end{center}
\end{figure}

The state-of-the-art NLO+NLL SUSY-QCD cross-section predictions for
the individual processes listed in Eq.\,(\ref{eq:processes}), occuring
in $p\bar{p}$ collisions at the Tevatron, are shown in
Fig.\,\ref{fig:xs_tev} as a function of the average mass $m$ of the
final state sparticles. For illustration we show these results for the
case $m_{\tilde{q}} = m_{\tilde{g}}$. In Fig.\,\ref{fig:xs_sum_tev}
the total NLO+NLL cross section for the sum of all four processes,
i.e.\ $\tilde{q}\tilde{q}\,, \tilde{q}\bar{\tilde{q}}\,,
\tilde{q}\tilde{g}$ and $ \tilde{g}\tilde{g}$ production, is
presented.  For the central values, the renormalization and
factorization scales are taken as $\mu = \mu_0 = m$, which is the
scale choice adopted as the preferred one in the context of the NLO
SUSY-QCD
calculations.\cite{Beenakker:1994an}\cdash\cite{Beenakker:1997ut} The
error band represents an estimate of the theoretical uncertainty of
the cross section prediction. It consists of the 68\% C.L.\ pdf and
$\alpha_{\rm s}$ error, added in quadrature, and the error from scale
variation in the range $m/2\le \mu \le 2m$ added linearly to the
combined pdf and $\alpha_{\rm s}$ uncertainty. By this linear
combination of scale uncertainty and combined pdf and $\alpha_{\rm s}$
errors we provide a conservative estimate of the theory error.

\begin{figure}
\begin{center}
\hspace*{-5mm}\epsfig{file=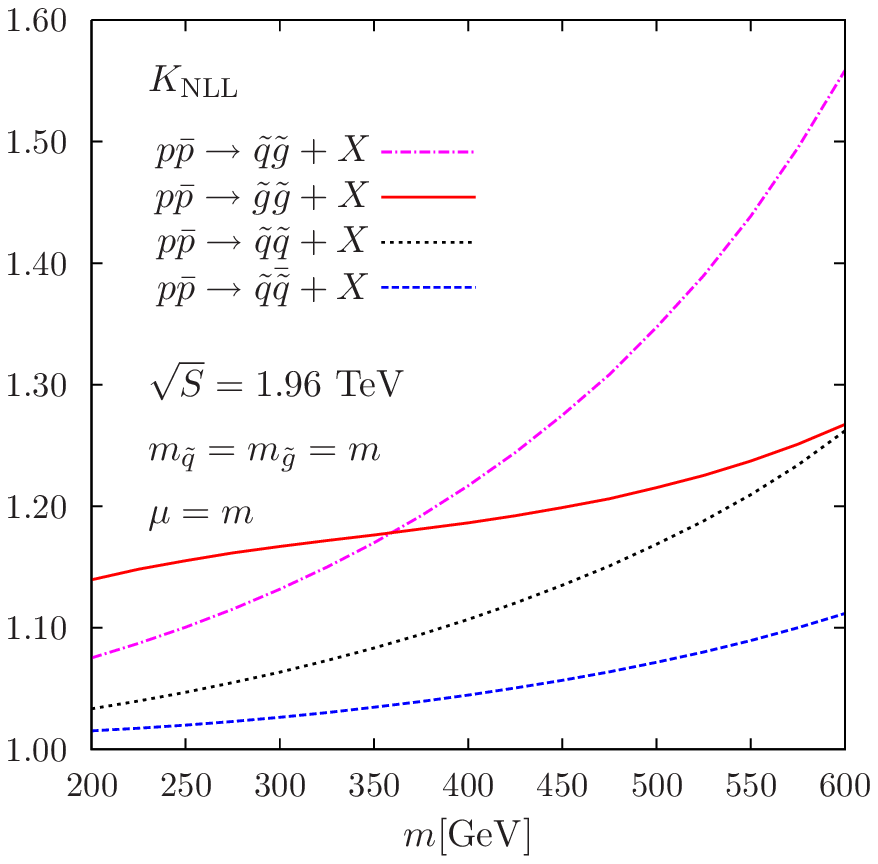, width=0.48\columnwidth,angle=0}
\hspace*{5mm}\epsfig{file=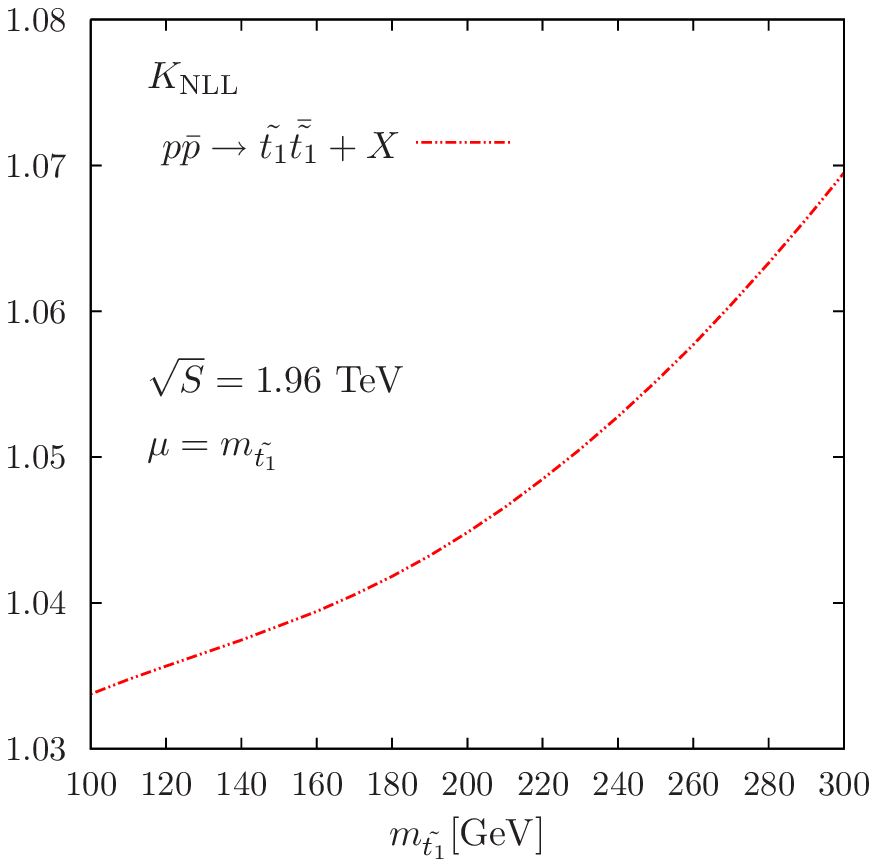, width=0.48\columnwidth,angle=0}
\caption{The NLL $K$-factor $K_{\rm
  NLL}=\sigma_{\rm NLO+NLL}/\sigma_{\rm NLO}$ for the individual
squark and gluino pair-production processes at the Tevatron as a function of the 
average sparticle mass $m$.
\label{fig:knll_tev}}
\end{center}
\end{figure}

At the scale $\mu =m$ the cross-section predictions are enhanced by
soft-gluon resummation. The relative $K$-factor $K_{\rm
  NLL}=\sigma_{\rm NLO+NLL}/\sigma_{\rm NLO}$ for this scale choice is
displayed in Fig.\,\ref{fig:knll_tev} for squark and gluino masses in
the range between 200 and 600\,GeV, and stop masses between 100 and
300\,GeV. The soft-gluon corrections are moderate for
$\tilde{q}\bar{\tilde{q}}$ and $\tilde t_1\bar{\tilde t}_1$
production, but increase the predictions for $\tilde{q}\tilde{q}$,
$\tilde{g}\tilde{g}$ and $\tilde{q}\tilde{g}$ final states by around
15, 20 and 40\%, respectively, assuming squark and gluino masses near
500\,GeV. Because of the increasing importance of the threshold
region, the corrections in general become larger for increasing
sparticle masses. The large effect of soft-gluon resummation for
$\tilde{q}\tilde{g}$ and $\tilde{g}\tilde{g}$ production can be mostly
attributed to the importance of gluon initial states for these
processes. Furthermore, the presence of gluinos in the final state
results in an enhancement of the NLL contributions, since in this case
the Casimir invariants that enter the $D_I$ coefficients in
Eq.~(\ref{eq:2}) are larger than for processes involving only
squarks. The substantial value of $K_{\rm NLL}$ for
$\tilde{q}\tilde{q}$ production at the Tevatron is a consequence of
the behaviour of the corresponding NLO
corrections\cite{Beenakker:1996ch}, which strongly decrease with
increasing squark mass.

\begin{figure}[h]
\begin{center}
\epsfig{file=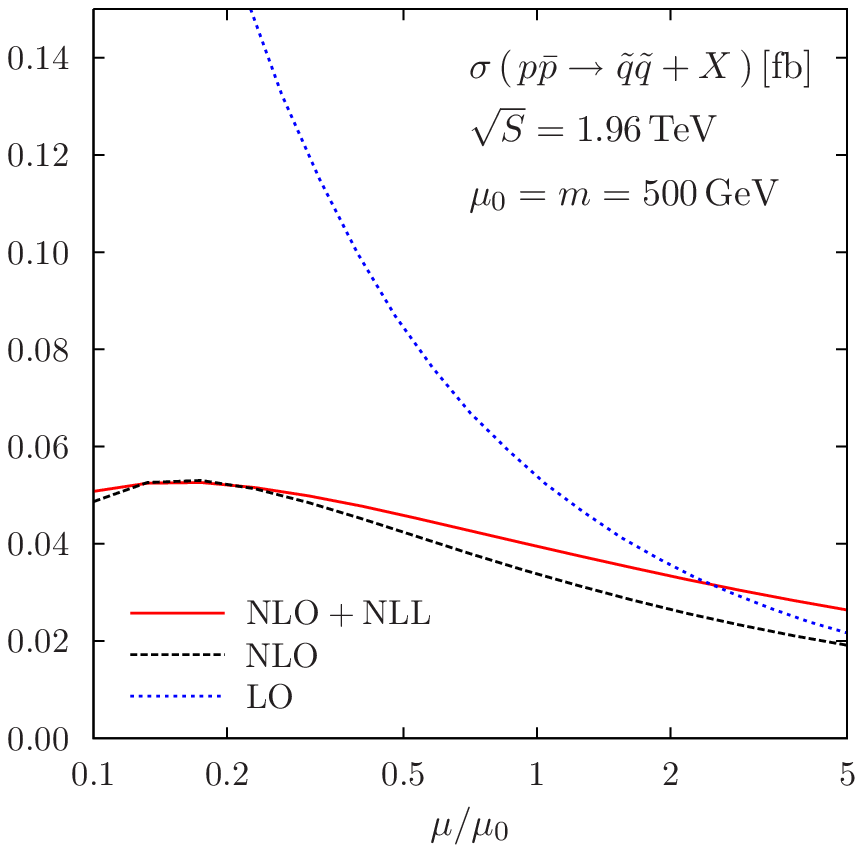, width=0.45\columnwidth,angle=0}
\epsfig{file=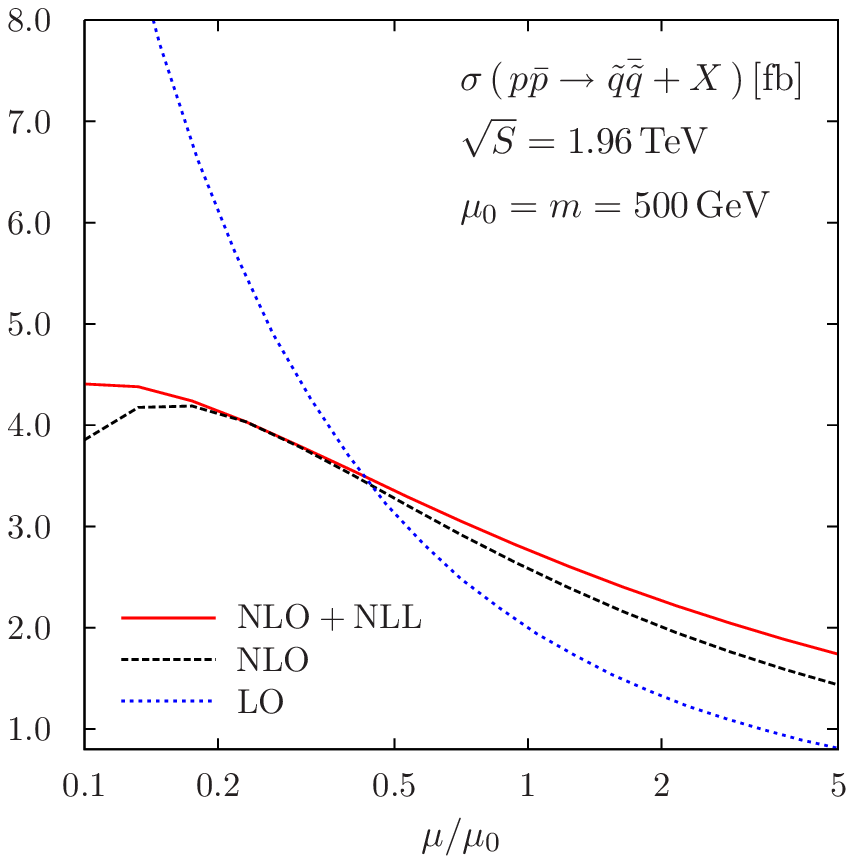, width=0.45\columnwidth,angle=0}\\[5mm]
\epsfig{file=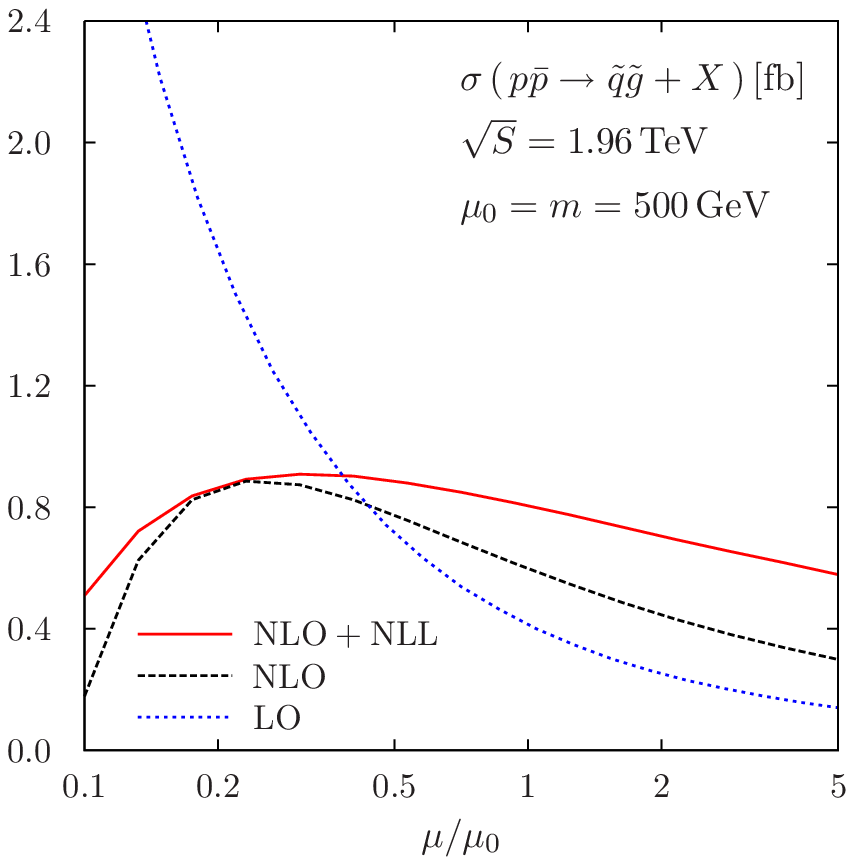, width=0.45\columnwidth, angle=0}
\epsfig{file=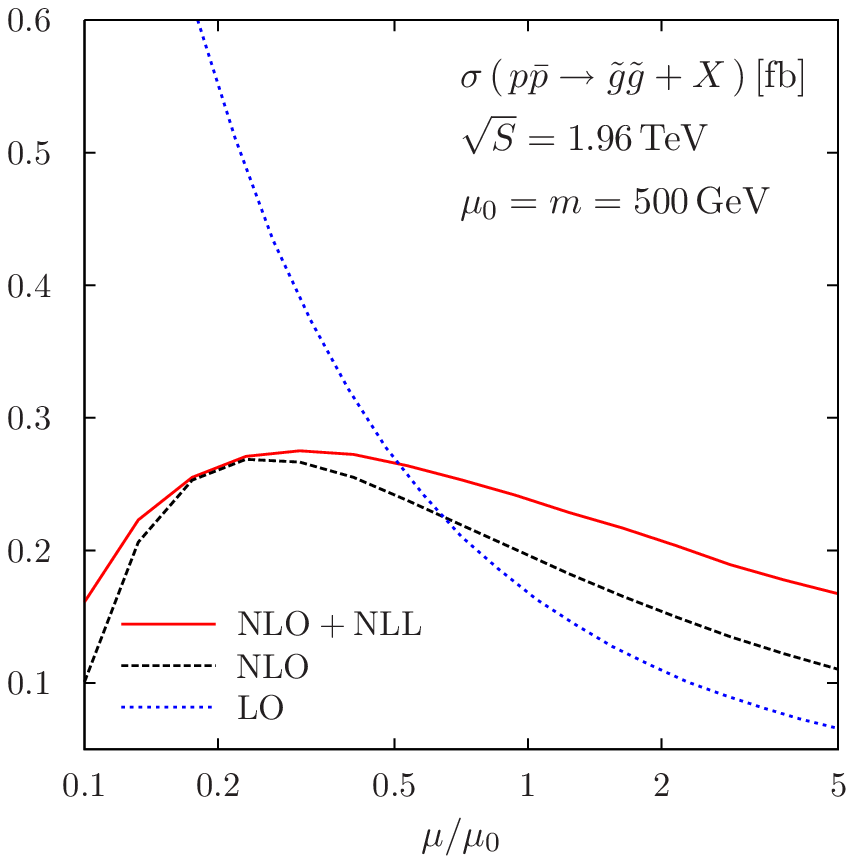, width=0.45\columnwidth, angle=0}\\[5mm]
\epsfig{file=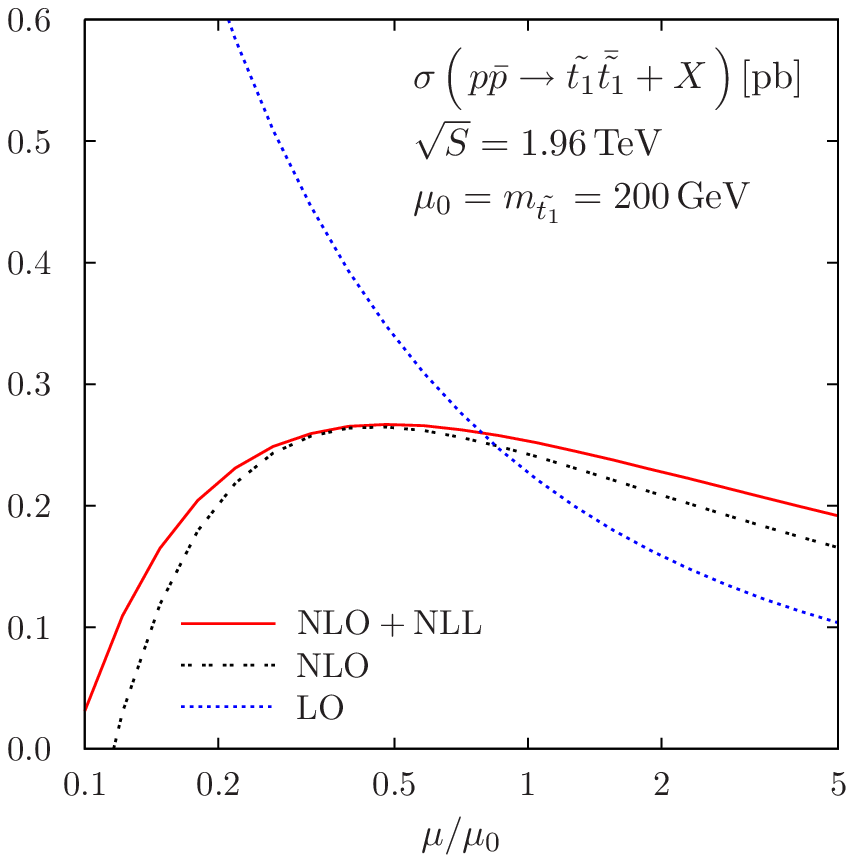, width=0.45\columnwidth, angle=0}
\end{center}
\caption{The scale dependence of the LO, NLO and NLO+NLL cross
  sections for the individual squark and gluino pair-production
  processes at the Tevatron.
The squark and gluino masses have been set equal $m_{\tilde{q}} = m_{\tilde{g}} = m$ in the upper four plots.  \label{fig:scale_tev}}
\end{figure}

Let us next discuss the scale dependence of the SUSY-QCD cross-section
prediction in some more detail.  Fig.\,\ref{fig:scale_tev} shows the
scale dependence in LO, NLO and NLO+NLL for the different production
processes, listed in Eq.\,(\ref{eq:processes}),
at the Tevatron. The mass of the $(\tilde{u}, \tilde{d}, \tilde{c},
\tilde{s}, \tilde{b})$-squarks and the gluino mass have been set to
$m_{\tilde{q}} = m_{\tilde{g}} = 500$\,GeV, while the stop mass is
fixed to $m_{\tilde{t}_1} = 200$\,GeV. Note that the LO predictions
are obtained with the LO MSTW pdf set\cite{MSTW:private} and the
corresponding LO value for $\alpha_{\rm s}$. The renormalization and
factorization scales are set equal to each other and varied around the
average mass of the final state sparticles, $\mu_0 = m$.  We observe
the usual strong reduction of the scale dependence when going from LO
to NLO. A further significant improvement is obtained when the
resummation of threshold logarithms is included, in particular for
$\tilde{g}\tilde{g}$ and $\tilde{q}\tilde{g}$ production. For the
$\tilde{q}\bar{\tilde{q}}\,, \tilde{q}\tilde{g}$ and
$\tilde{g}\tilde{g}$ final states, contributing the most to the
inclusive squark and gluino cross section at the Tevatron, the LO, NLO
and NLO+NLL cross section predictions converge particularly well near
$\mu = m/2$, which appears to be the preferred scale choice for these
processes.

\subsection{LHC @ 7\,TeV}
SUSY searches at the LHC, which is currently operating at 7\,TeV, will
soon be sensitive to sparticles with masses in the TeV-range. We thus
discuss the cross section predictions for the LHC @ 7\,TeV in more
detail, including in particular the dependence of the NLL corrections
on the ratio of the gluino and squark masses, and the size of the
theory uncertainty due to scale, pdf and $\alpha_{\rm s}$ errors.

\begin{figure}
\begin{center}
\epsfig{file=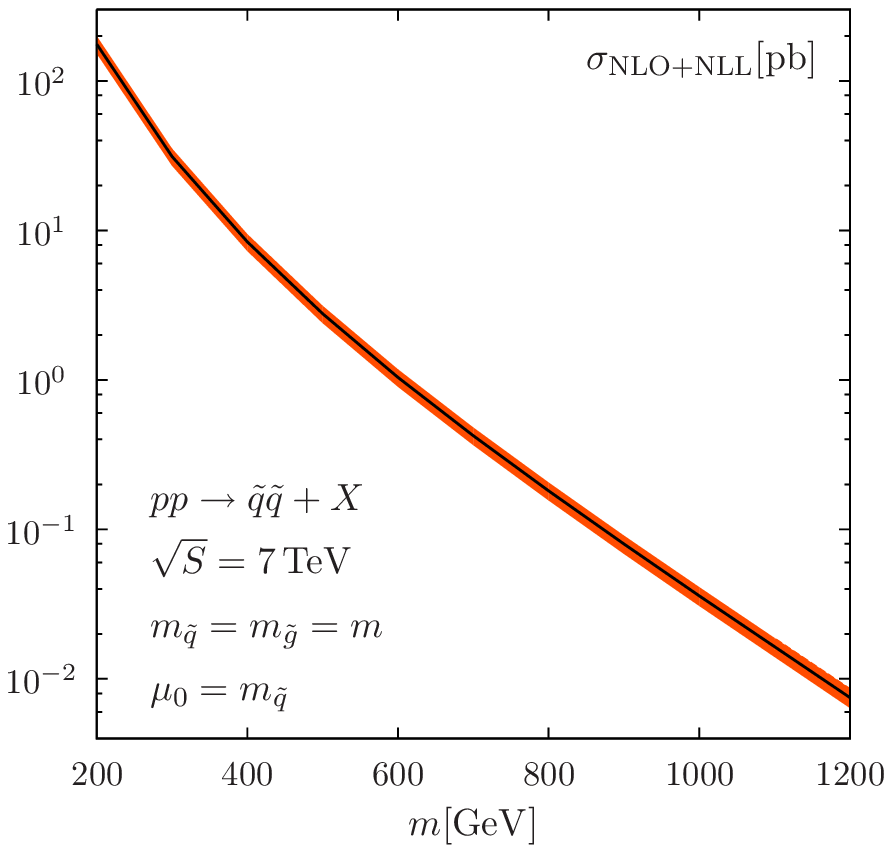,
  width=0.45\columnwidth,angle=0}
\epsfig{file=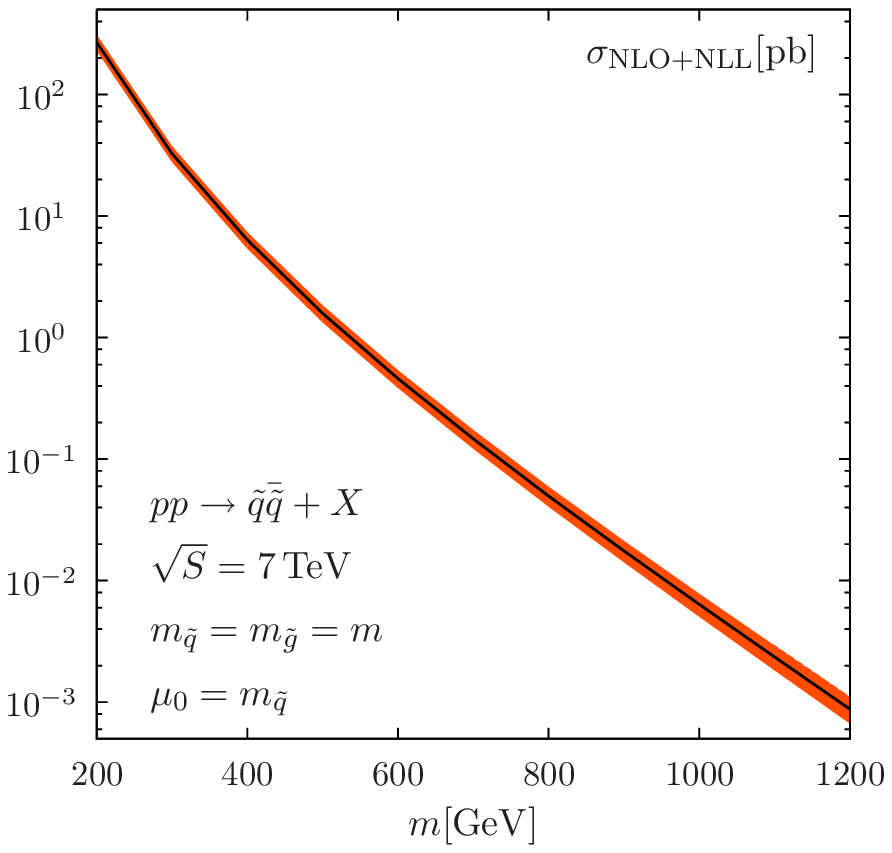,
  width=0.45\columnwidth,angle=0}\\[5mm]
\epsfig{file=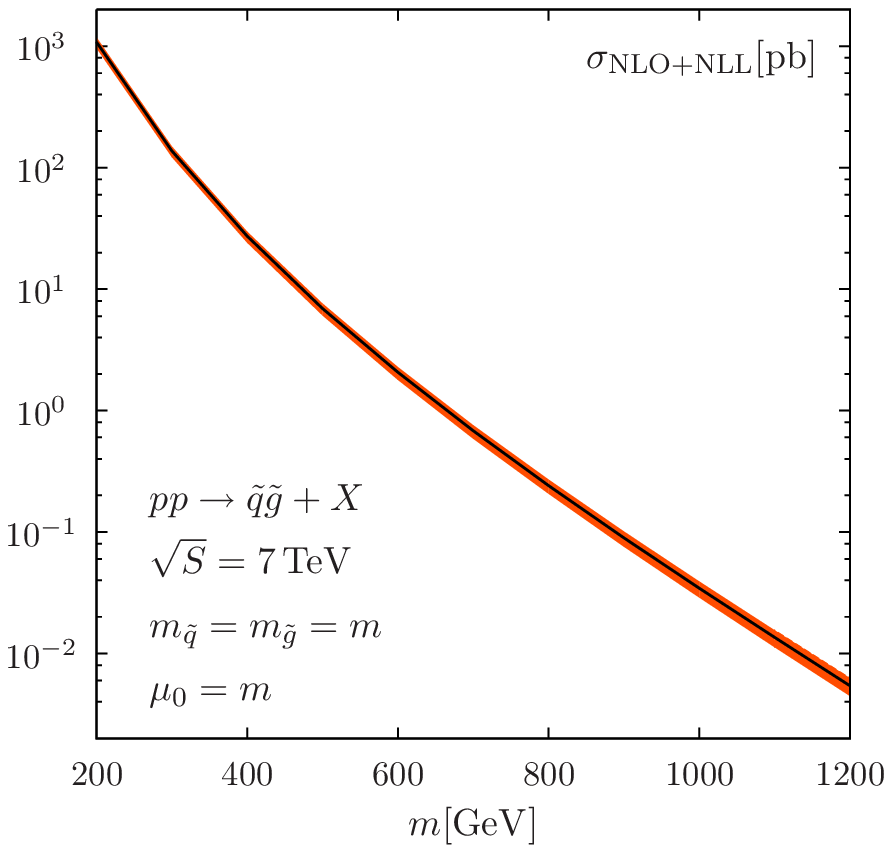,
  width=0.45\columnwidth,angle=0}
\epsfig{file=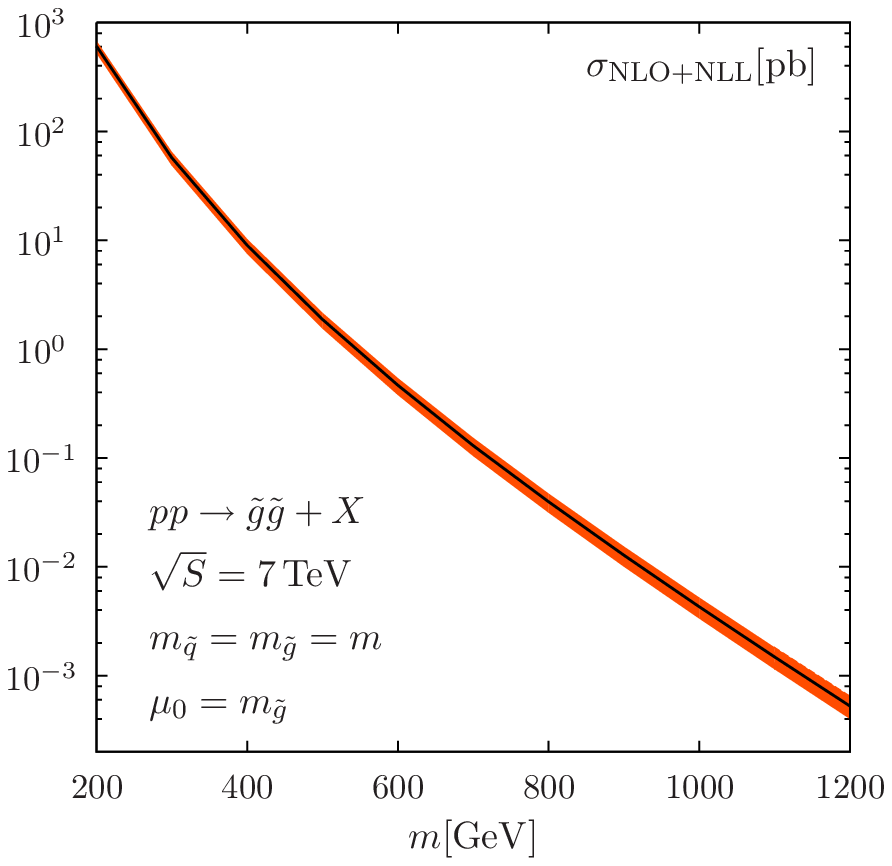,
  width=0.45\columnwidth,angle=0}\\[5mm]
\epsfig{file=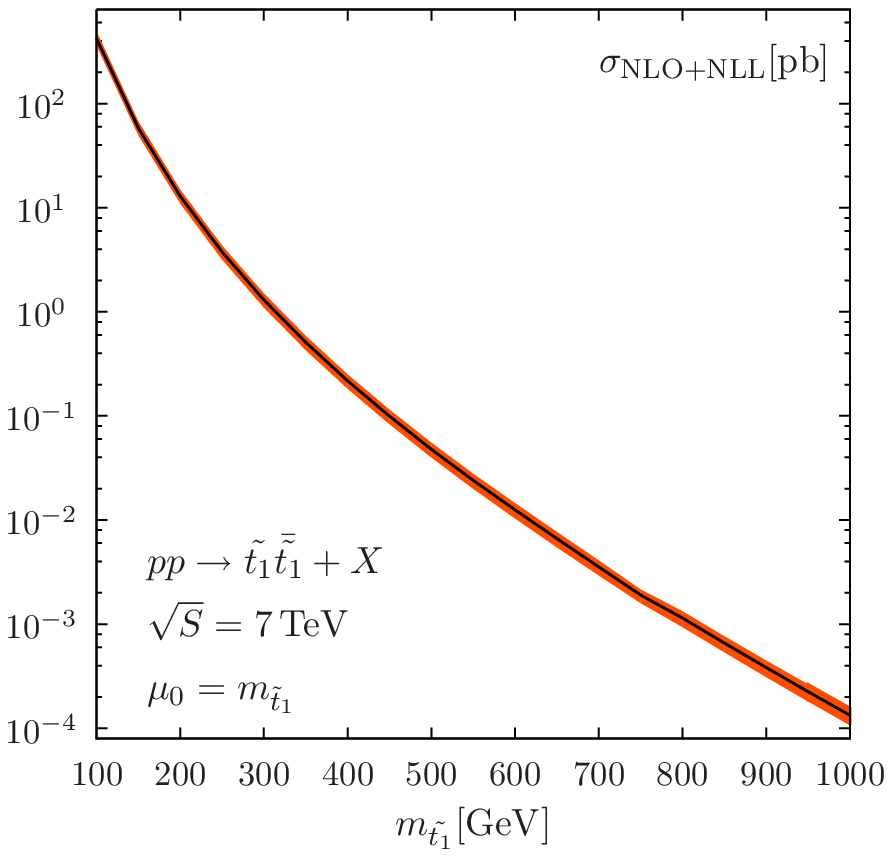,
  width=0.45\columnwidth,angle=0}
\end{center}
\caption{The NLO+NLL SUSY-QCD cross section for the individual squark
  and gluino pair-production processes at the LHC with 7\,TeV, $pp \to
  \tilde{q}\tilde{q}\,, \tilde{q}\bar{\tilde{q}}\,,
  \tilde{q}\tilde{g}\,, \tilde{g}\tilde{g}+X$ and $pp \to\tilde
  t_1\bar{\tilde t}_1+ X$, as a function of the average sparticle mass
  $m$. The error band includes the 68\% C.L.\ pdf and $\alpha_{\rm s}$
  error, added in quadrature, and the error from scale variation in
  the range $m/2\le \mu \le 2m$ added linearly to the combined pdf and
  $\alpha_{\rm s}$ uncertainty.\label{fig:xs_lhc7}}
\end{figure}

\begin{figure}
\begin{center}
\epsfig{file=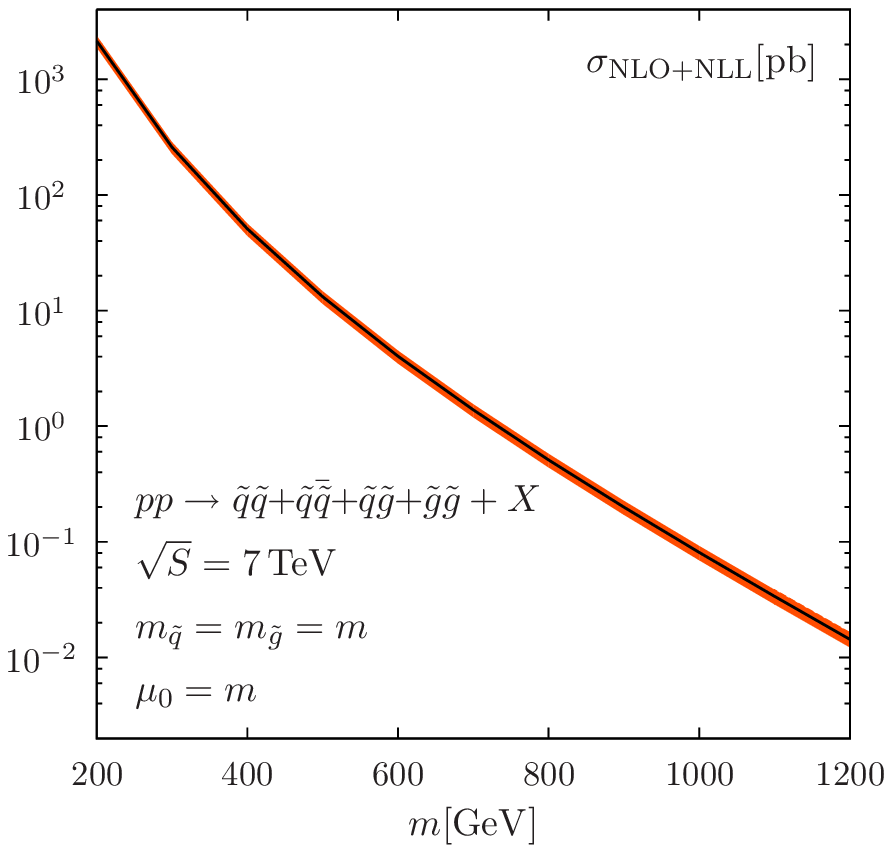,
  width=0.55\columnwidth,angle=0}
\caption{The NLO+NLL SUSY-QCD cross section for inclusive squark and
  gluino pair-production at the LHC with 7\,TeV, $pp \to
  \tilde{q}\tilde{q} + \tilde{q}\bar{\tilde{q}} + \tilde{q}\tilde{g}+
  \tilde{g}\tilde{g}+X$, as a function of the average sparticle mass
  $m$. The error band includes the 68\% C.L.\ pdf and $\alpha_{\rm s}$
  error, added in quadrature, and the error from scale variation in
  the range $m/2\le \mu \le 2m$ added linearly to the combined pdf and
  $\alpha_{\rm s}$ uncertainty.
\label{fig:xs_sum_lhc7}}
\end{center}
\end{figure}

As before, we first present the NLO+NLL cross section predictions for
the five individual processes and for the sum of the
$\tilde{q}\tilde{q}\,, \tilde{q}\bar{\tilde{q}}\,,
\tilde{q}\tilde{g}\,, \tilde{g}\tilde{g}$ final states, cf.\
Figs.\,\ref{fig:xs_lhc7} and \ref{fig:xs_sum_lhc7}. We include the
estimate of the theory uncertainty obtained from adding linearly the
scale dependence in the range $m/2\le \mu \le 2m$ to the combined 68\%
C.L.\ pdf and $\alpha_{\rm s}$ error, added in quadrature.

\begin{figure}
\begin{center}
\hspace*{-5mm}\epsfig{file=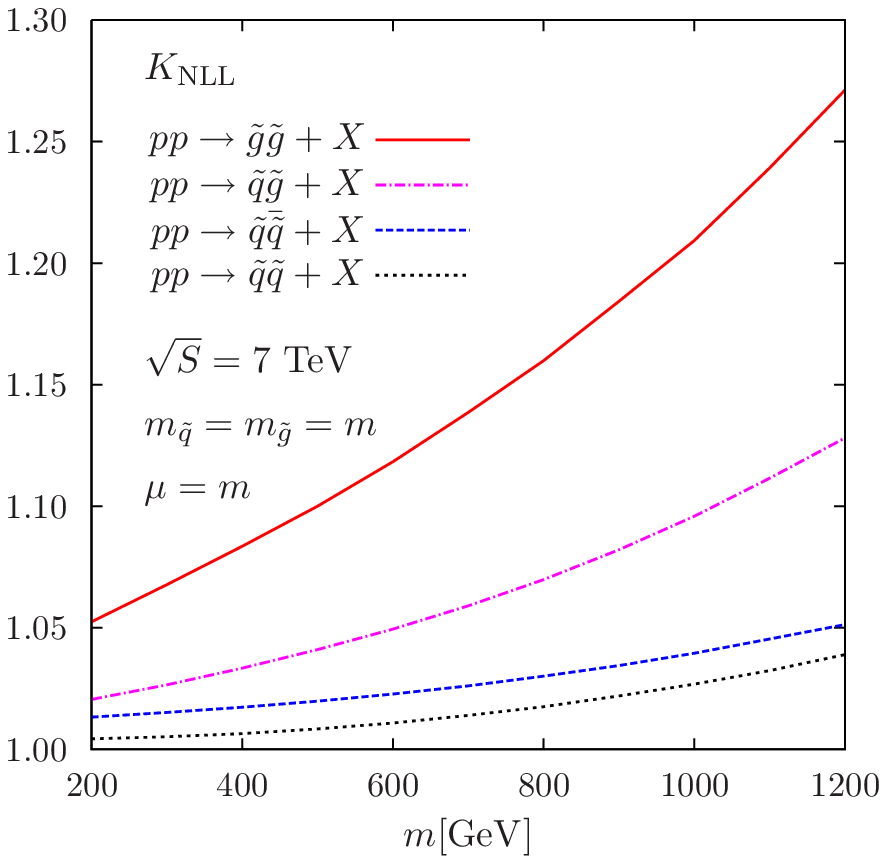, width=0.48\columnwidth,angle=0}
\hspace*{5mm}\epsfig{file=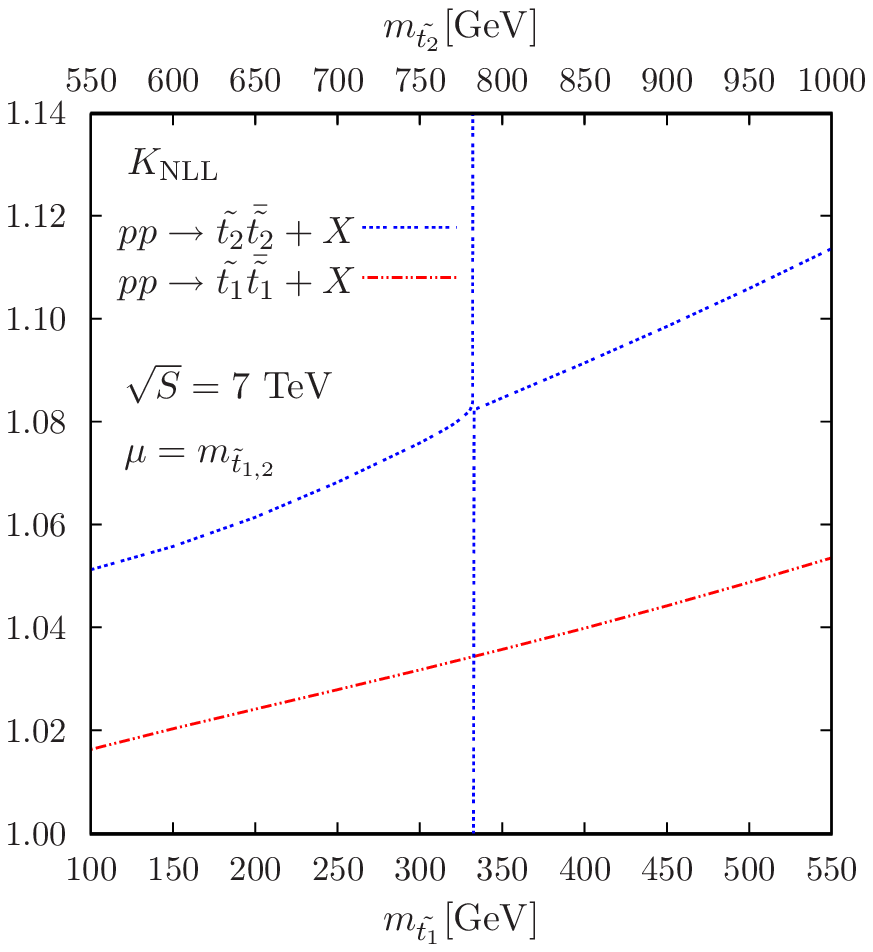, width=0.48\columnwidth,angle=0}
\caption{The NLL $K$-factor $K_{\rm
  NLL}=\sigma_{\rm NLO+NLL}/\sigma_{\rm NLO}$ for the individual squark and gluino
pair-production processes at the LHC with 7\,TeV as a function of the average sparticle mass $m$.
\label{fig:knll_lhc7}}
\end{center}
\end{figure}

\begin{figure}
\begin{center}
\hspace*{5mm}\epsfig{file=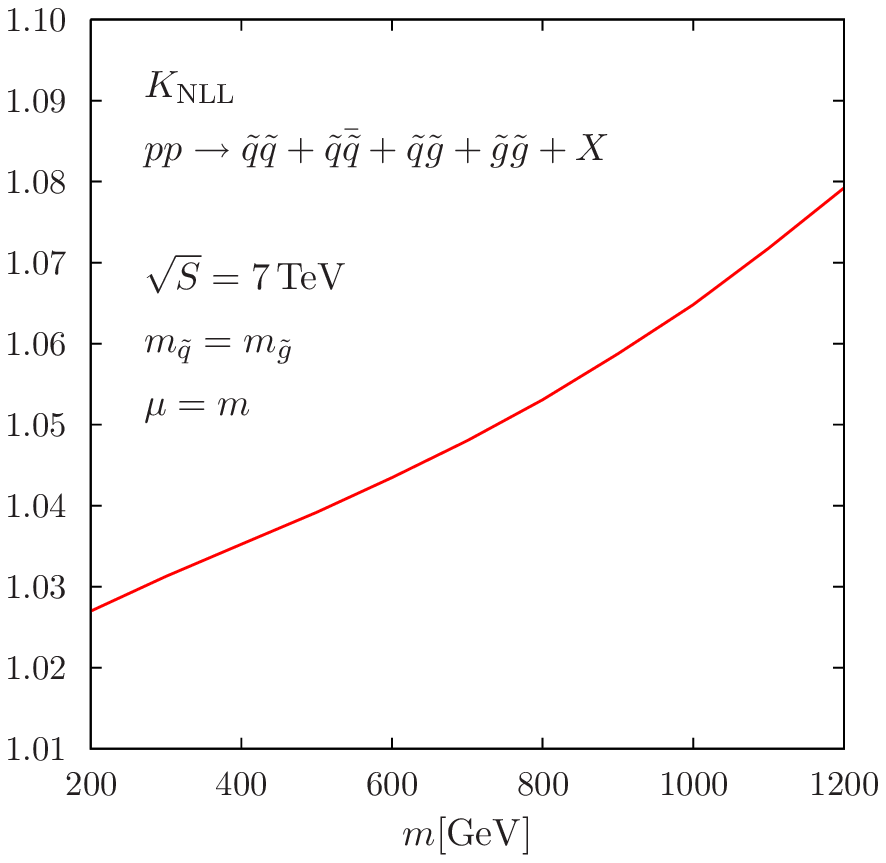, width=0.48\columnwidth,angle=0}
\caption{The NLL $K$-factor $K_{\rm
  NLL}=\sigma_{\rm NLO+NLL}/\sigma_{\rm NLO}$ for for inclusive squark and gluino
  pair-production at the LHC with 7\,TeV, $pp \to \tilde{q}\tilde{q} +
  \tilde{q}\bar{\tilde{q}} + \tilde{q}\tilde{g}+
  \tilde{g}\tilde{g}+X$, as a function of the average sparticle mass
  $m$.
\label{fig:knll_inc_lhc7}}
\end{center}
\end{figure}

In Fig.\,\ref{fig:knll_lhc7} we show the enhancement of the cross
section due to the NLL resummation at the scale $\mu =m$. For
gluino-pair and squark-gluino production processes we find a
significant increase of 10-20\% at masses around 1\,TeV. An
enhancement of up to 10\% can also be observed for the production of
heavy stop particles. Note that the singularities at the stop-decay
threshold $m_{\tilde{t}}=m_t+m_{\tilde{g}} = 782.5$\,GeV originate
from the stop wave-function renormalization. They are an unphysical
artefact of an on-shell scheme approach\cite{Beenakker:1997ut} and
could be removed by taking into account the finite widths of the
unstable stops. The effect of the NLL resummation on the cross section
for inclusive squark and gluino production is shown in
Fig.\,\ref{fig:knll_inc_lhc7}.

\begin{figure}
\begin{center}
\epsfig{file=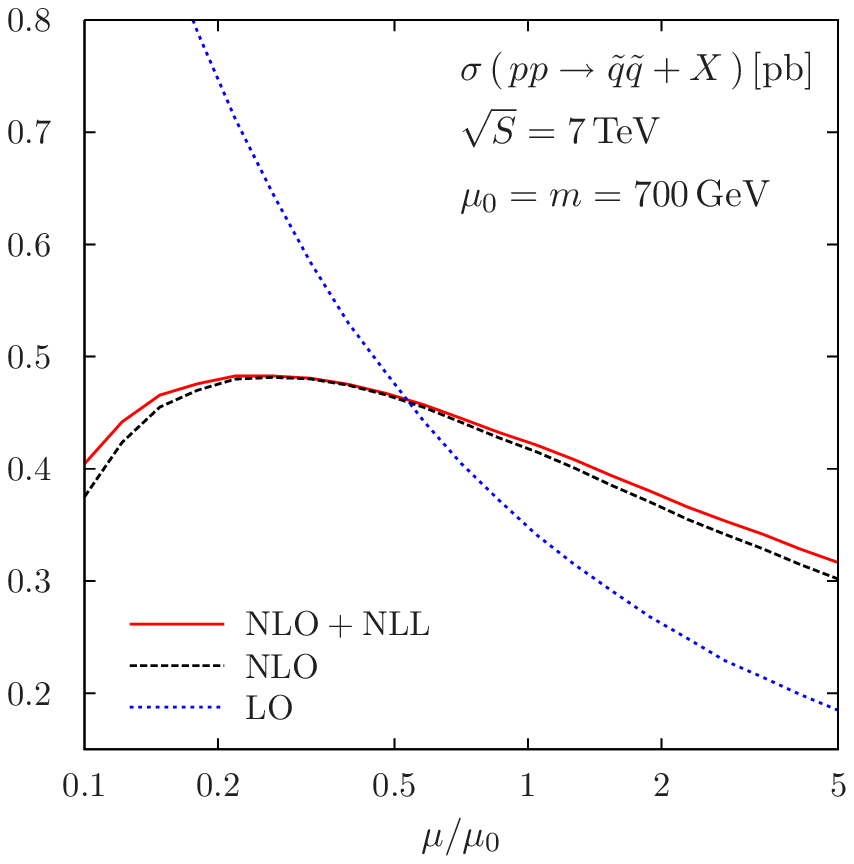, width=0.45\columnwidth,angle=0}
\epsfig{file=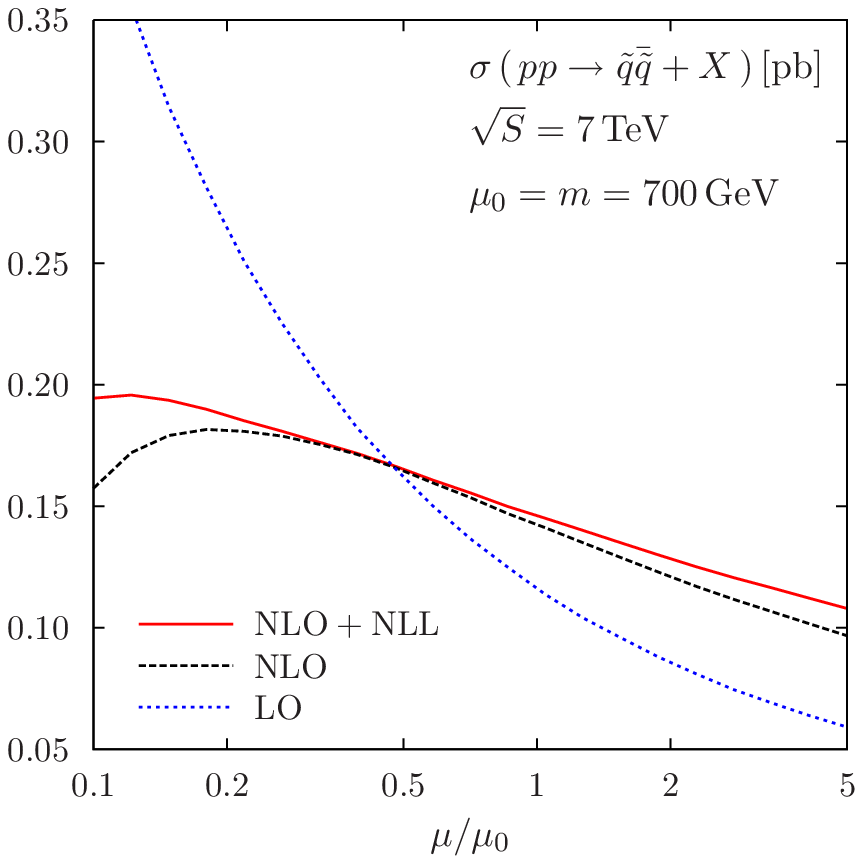, width=0.45\columnwidth,angle=0}\\[5mm]
\epsfig{file=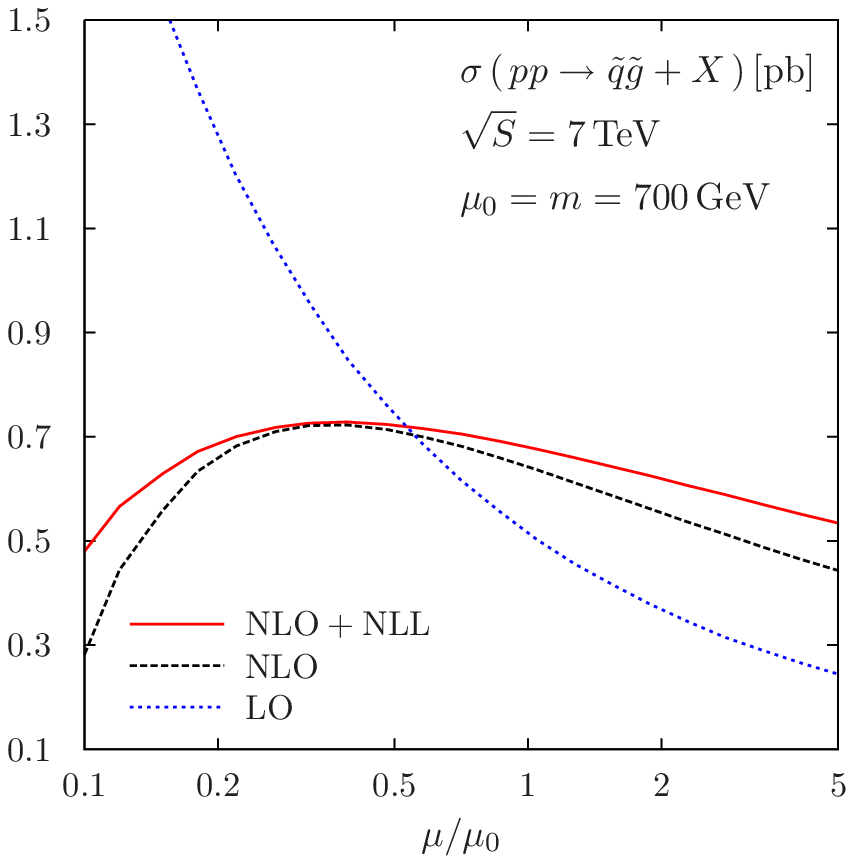, width=0.45\columnwidth, angle=0}
\epsfig{file=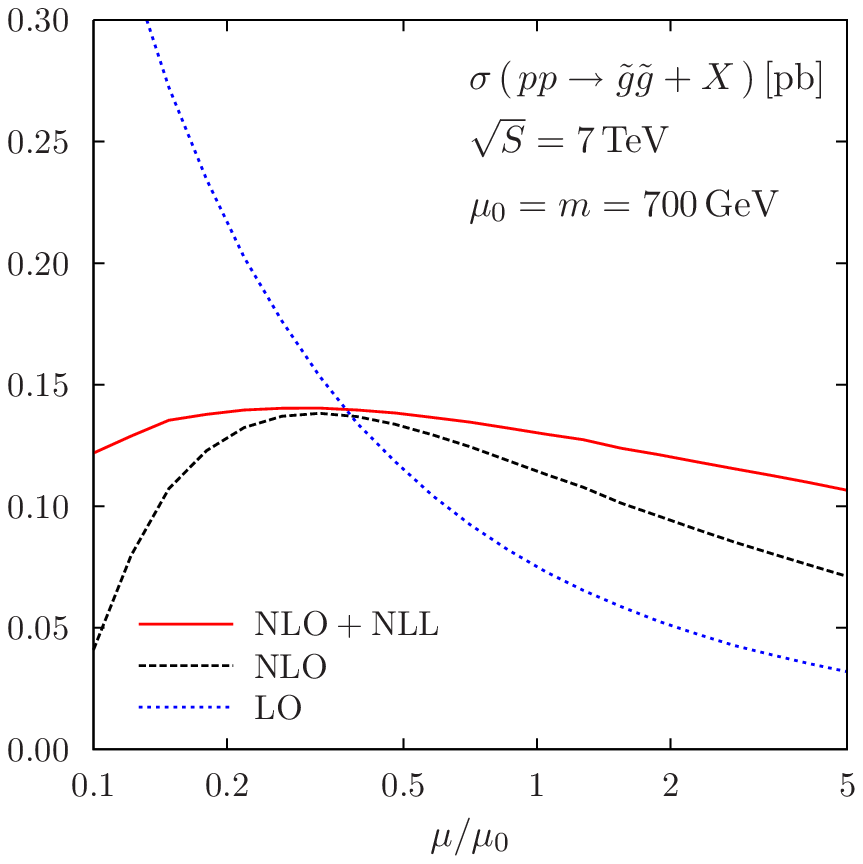, width=0.45\columnwidth, angle=0}\\[5mm]
\epsfig{file=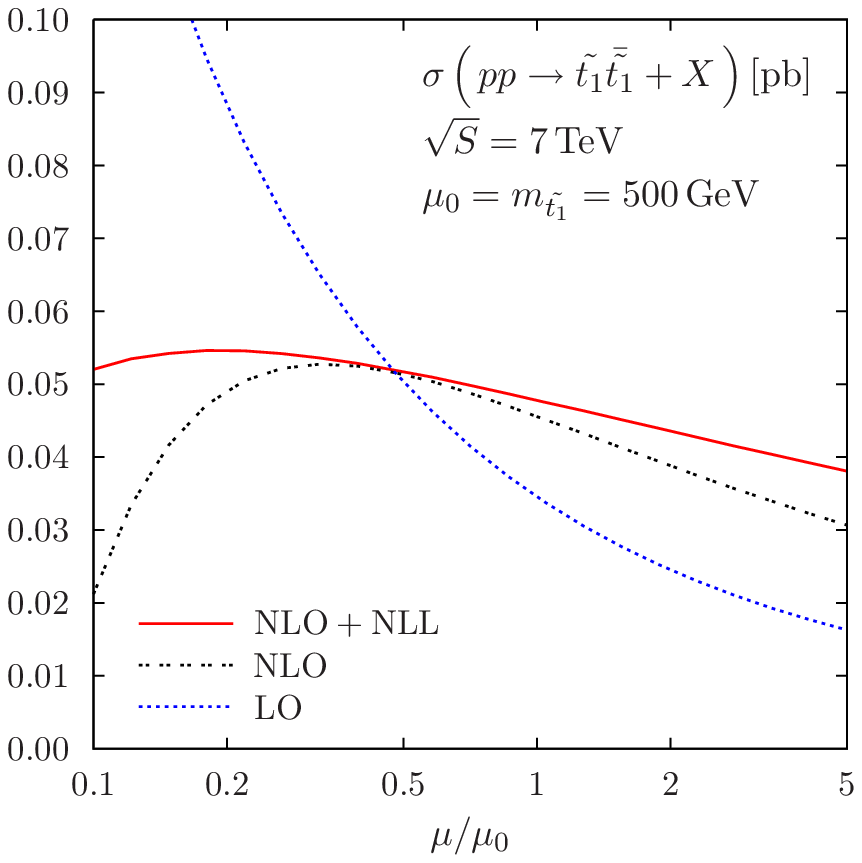, width=0.45\columnwidth, angle=0}
\end{center}
\caption{The scale dependence of the LO, NLO and NLO+NLL cross
  sections for the individual squark and gluino pair-production
  processes at the LHC with 7\,TeV.
  The squark and gluino masses have been set equal $m_{\tilde{q}} = m_{\tilde{g}} = m$ in the upper four plots. 
  \label{fig:scale_lhc7}}
\end{figure}

Next, we present the scale dependence in LO, NLO and NLO+NLL for the
various production processes, see Fig.\,\ref{fig:scale_lhc7}. Here,
the mass of the $(\tilde{u}, \tilde{d}, \tilde{c}, \tilde{s},
\tilde{b})$-squarks and the gluino mass have been set to
$m_{\tilde{q}} = m_{\tilde{g}} = 700$\,GeV, while the stop mass is
fixed to $m_{\tilde{t}_1} = 500$\,GeV. As anticipated, we observe a
significant reduction of the scale dependence when going from LO to
NLO and from NLO to NLO+NLL. Note that also for squark and gluino
production at the LHC, the convergence of the LO, NLO and NLO+NLL
cross section predictions near the scale $\mu = m/2$ is striking.

As the cross section predictions for the LHC operating at 7\,TeV are
of particular phenomenological importance, let us discuss the results
in some more detail. First, we collect representative values for LO,
NLO and NLO+NLL cross sections in Tables\,\ref{tbl1}-\ref{tbl3}. We
include the scale dependence in the range $m/2\le \mu \le 2m$ for the
LO, NLO and NLO+NLL calculations, and the 68\% C.L.\ pdf \footnote{The
  pdf errors presented in the tables apply to individual processes
  only. In general, the pdf error for inclusive squark and gluino
  production will differ from the one obtained through combination of
  errors for individual processes due to possible correlations between
  various production channels.} and $\alpha_{\rm s}$ error at NLO. The
NLO and NLL $K$-factors are displayed for convenience. The NLO and
NLO+NLL theory predictions with the default MSTW pdf
set\cite{MSTW:private} are compared to those obtained with the CT10
pdf set\cite{Lai:2010vv}. The LO cross sections are calculated with
the LO MSTW\cite{MSTW:private} and the CTEQ6L1\cite{CTEQ6LO}
pdfs. While the LO cross sections differ significantly between the two
pdf sets, the central NLO+NLL predictions are consistent within the
theoretical uncertainty.  Note, however, that the 68\% C.L.\ pdf error
estimate for the CT10 set, obtained through rescaling of the 90\% CL
error estimate\cite{Botje:2011sn}, is significantly larger than that
of MSTW.

\begin{figure}
\begin{center}
\epsfig{file=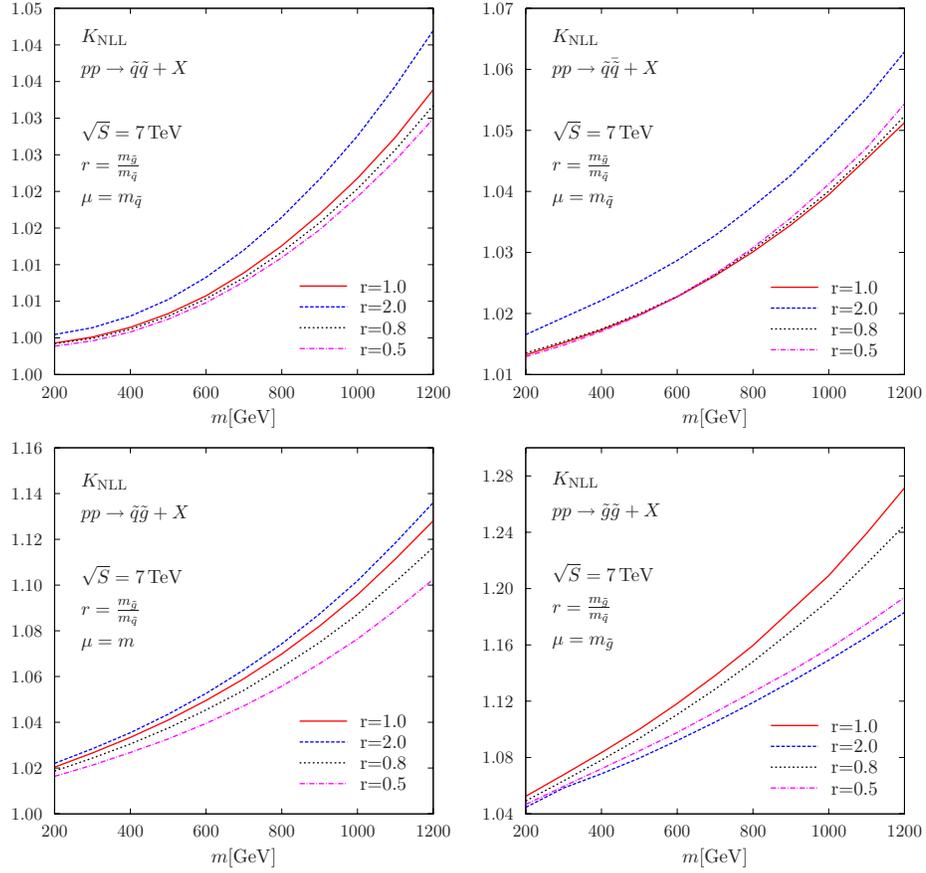, width=0.95\columnwidth,angle=0}\\
\end{center}
\caption{The relative NLL $K$-factor $K_{\rm NLL} =\sigma_{\rm
    NLO+NLL}/\sigma_{\rm NLO}$ for the individual squark and gluino
  pair-production processes as a function of the average sparticle
  mass $m$, for various mass ratios $r =
  m_{\tilde{g}}/m_{\tilde{q}}$.\label{fig:knll_r_lhc7}}
\end{figure}

\begin{figure}
\begin{center}
\epsfig{file=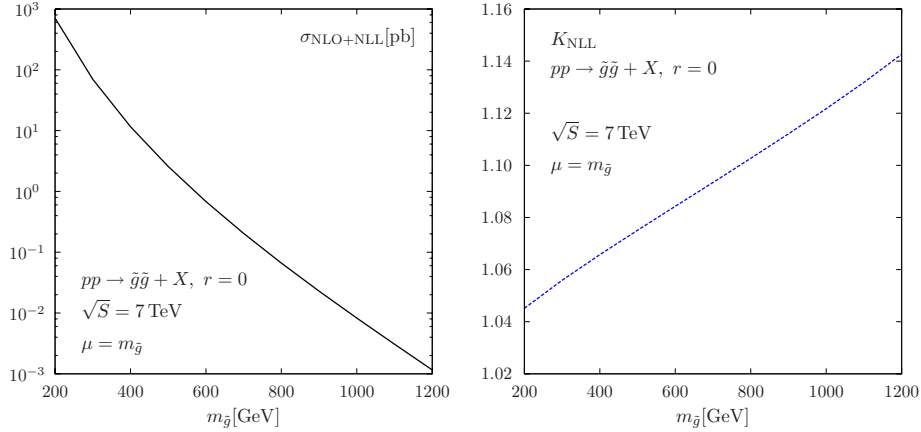, width=0.95\columnwidth,angle=0}\\
\end{center}
\caption{The NLO+NLL SUSY-QCD cross section (left) and NLL $K$-factor
  (right) for gluino pair-production in the heavy squark limit, $r =
  m_{\tilde{g}}/m_{\tilde{q}} = 0$, at the LHC with 7\,TeV, as a
  function of the gluino mass $m_{\tilde{g}}$.\label{fig:r0_lhc7}}
\end{figure}

The size of the cross section and the impact of the higher-order
corrections depends on the ratio $r = m_{\tilde{g}}/m_{\tilde{q}}$.
In Fig.\,\ref{fig:knll_r_lhc7} we thus present the NLL $K$-factor for
different values of $r$ for individual processes with the
$\tilde{q}\tilde{q}\,, \tilde{q}\bar{\tilde{q}}\,,
\tilde{q}\tilde{g}\,, \tilde{g}\tilde{g}$ final states . In general,
in the range of values of $r$ studied here, the dependence of the
soft-gluon enhancement on $r$ is moderate.  Of particular interest is
the limit $r\to 0$, i.e.\ the limit of very heavy squarks.  Such
scenarios are predicted within so-called models of split
supersymmetry\cite{ArkaniHamed:2004fb} and lead to interesting new
phenomenological signatures at the LHC, like long-lived
gluinos.\cite{Khachatryan:2010uf,Aad:2011yf} In
Fig.\,\ref{fig:r0_lhc7} we present the NLO+NLL cross section
prediction for gluino-pair production in the limit $r = 0$ and show
the impact of soft-gluon resummation. The $\tilde{g}\tilde{g}$ cross
section for $r=0$ is higher than for $r=1$, with the difference
growing from a factor of 1.2 at $m=200$ GeV up to a factor of 4.7 for
$m=1.2$ TeV. The impact of soft-gluon resummation is significant, and
the NLL corrections enhance the cross section by about 12\% for
gluinos with masses around 1\,TeV.

\begin{figure}
\begin{center}
\epsfig{file=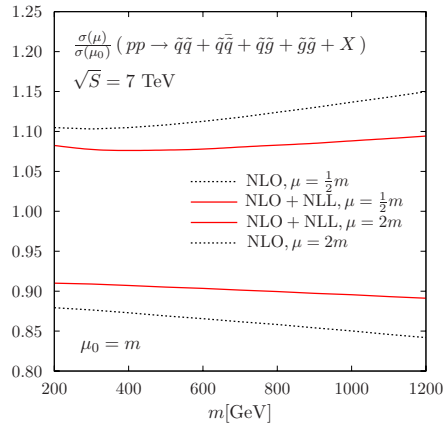, width=0.45\columnwidth,angle=0}\\
\end{center}
\caption{The scale dependence of the NLO and NLO+NLL SUSY-QCD cross
  sections for inclusive squark and gluino pair-production at the LHC
  with 7\,TeV, $pp \to \tilde{q}\tilde{q} + \tilde{q}\bar{\tilde{q}} +
  \tilde{q}\tilde{g}+ \tilde{g}\tilde{g}+X$, as a function of the
  average sparticle mass $m$. Shown are results for the mass ratio $r
  = m_{\tilde{g}}/m_{\tilde{q}} = 1$.  The upper two curves correspond
  to the scale set to $\mu = m/2$, the lower two curves to $\mu =
  2m$.\label{fig:scale_mass_lhc7}}
\end{figure}

\begin{figure}
\begin{center}
\epsfig{file=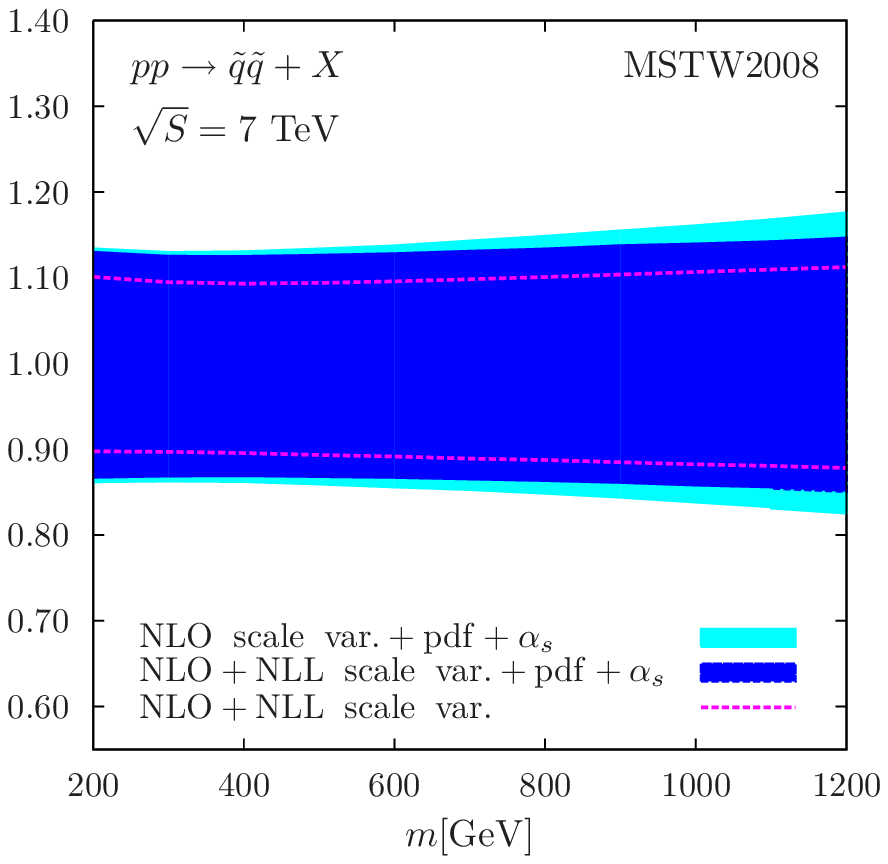, width=0.45\columnwidth,angle=0}
\epsfig{file=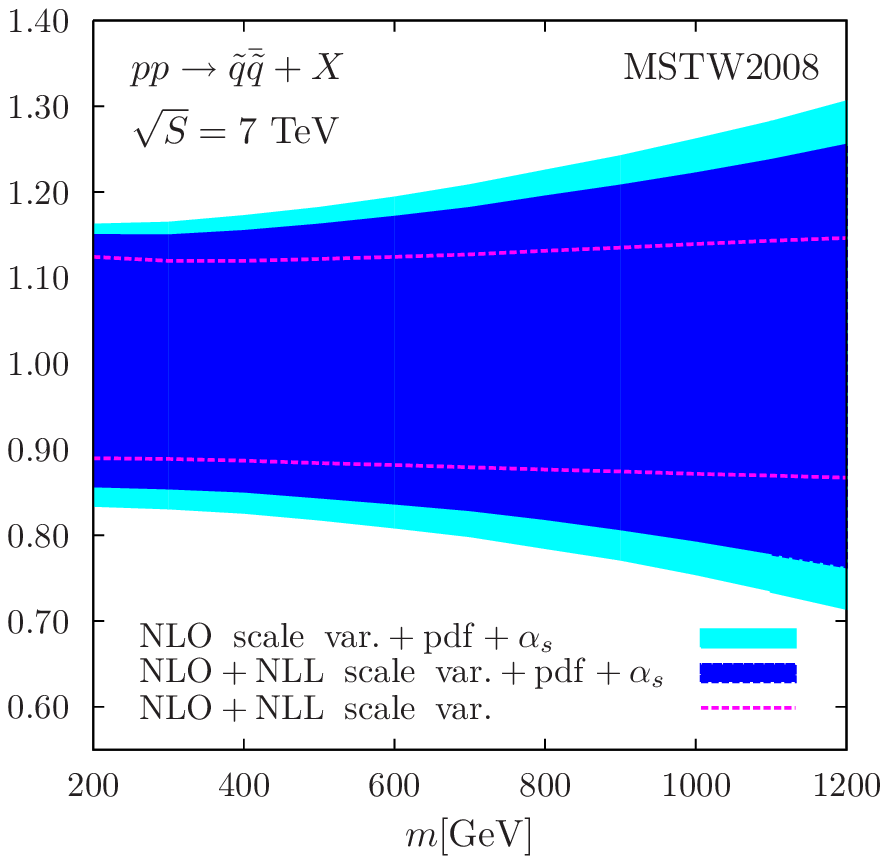, width=0.45\columnwidth,angle=0}\\[5mm]
\epsfig{file=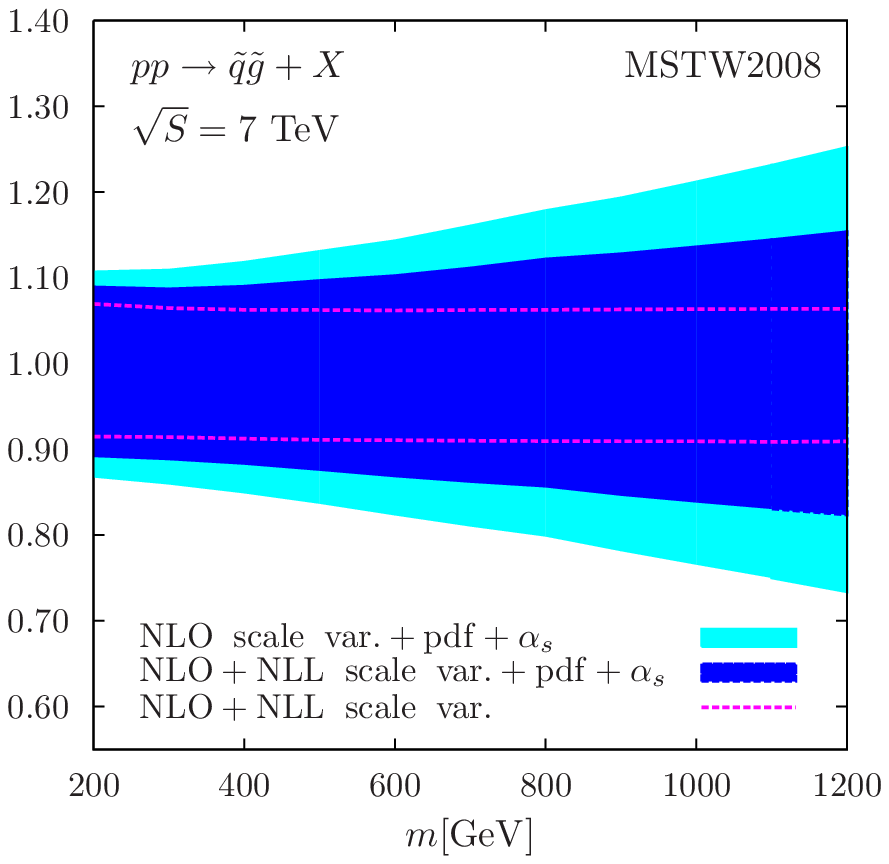, width=0.45\columnwidth,angle=0}
\epsfig{file=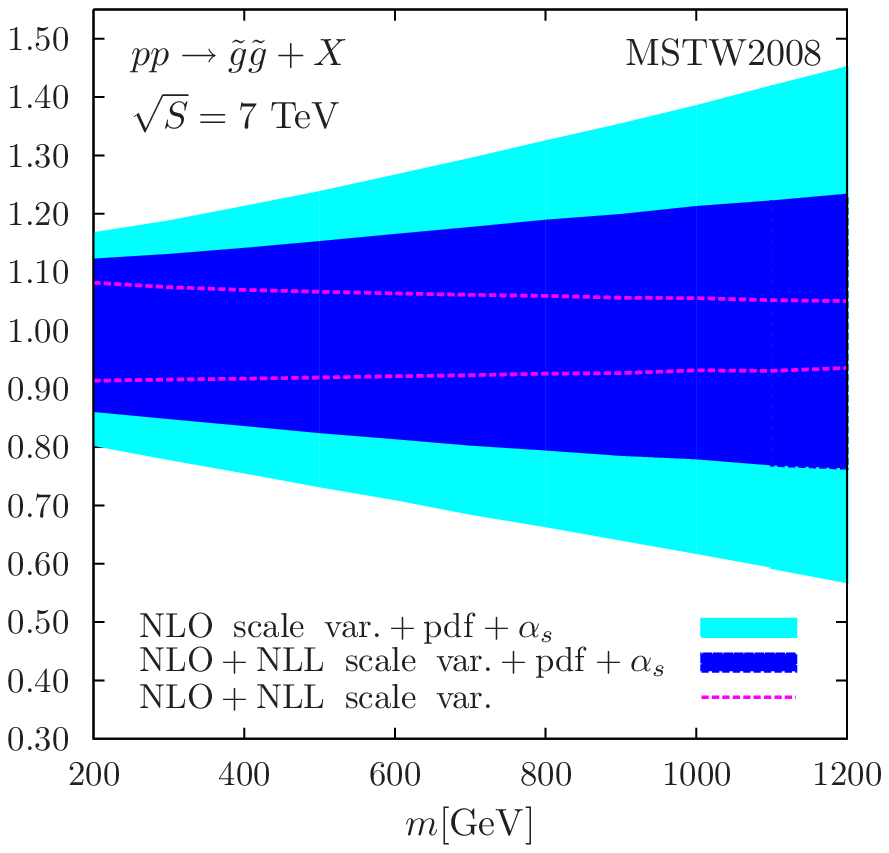, width=0.45\columnwidth,angle=0}
\end{center}
\caption{The theoretical uncertainty for the the individual squark and
  gluino pair-production processes at the LHC with 7\,TeV, $pp \to
  \tilde{q}\tilde{q}\,, \tilde{q}\bar{\tilde{q}}\,,
  \tilde{q}\tilde{g}\,, \tilde{g}\tilde{g}+X$, as a function of the
  sparticle mass $m_{\tilde{q}} = m_{\tilde{g}} = m$. The error bands represent the NLO+NLL
  scale uncertainty in the the range $m/2\le \mu \le 2m$, and the
  total theory uncertainty including the 68\% C.L.\ pdf and
  $\alpha_{\rm s}$ error, added in quadrature, and the error from
  scale variation in the range $m/2\le \mu \le 2m$ added linearly to
  the combined pdf and $\alpha_{\rm s}$ uncertainty. The total theory
  uncertainty is shown at both NLO and
  NLO+NLL.\label{fig:scale_mass_ind_lhc7}}
\end{figure}

The enhancement of the cross section prediction due to higher-order
corrections depends, of course, on the choice of scale. Choosing $\mu
= m/2$, we find in general NLL $K$-factors close to one for sparticle
cross sections at the LHC with 7\,TeV. However, a crucial improvement
when going from LO to NLO and from NLO to NLO+NLL is the reduction of
the scale dependence. Fig.\,\ref{fig:scale_mass_lhc7} shows a
comparison of the NLO and the NLO+NLL scale dependence for the sum of
the $\tilde{q}\tilde{q}\,, \tilde{q}\bar{\tilde{q}}\,,
\tilde{q}\tilde{g}\,, \tilde{g}\tilde{g}$ final states as a function
of the average sparticle mass, assuming $r = 1$. Threshold resummation
leads to a significant reduction of the scale dependence over the full
range of sparticle masses, with an overall scale uncertainty at
NLO+NLL of less than 10\%. In a phenomenological analysis, the
different production channels may contribute with different weight,
depending in detail on the signature and the choice of selection
cuts. Thus, we plot the scale uncertainty at NLO+NLL for the different
channels in Fig.\,\ref{fig:scale_mass_ind_lhc7}.  We also show the
full theory uncertainty, consisting of the 68\% C.L.\ pdf and
$\alpha_{\rm s}$ error added in quadrature, combined linearly with the
scale variation error for the NLO cross sections and for the NLO+NLL
cross sections.  We find that even though the pdf uncertainty is
significant, the inclusion of threshold resummation leads to a
sizeable reduction of the overall theory uncertainty. This is
particularly true for the case of gluino-pair and squark-gluino
production. For gluino-pair production, the total theory uncertainty
can be reduced by as much as a factor of two when going from NLO to
NLO+NLL. Looking at all different production processes, the overall
theory uncertainty at NLO+NLL is approximately 20\% or smaller.

\subsection{LHC @ 14 TeV}
\begin{figure}
\begin{center}
\epsfig{file=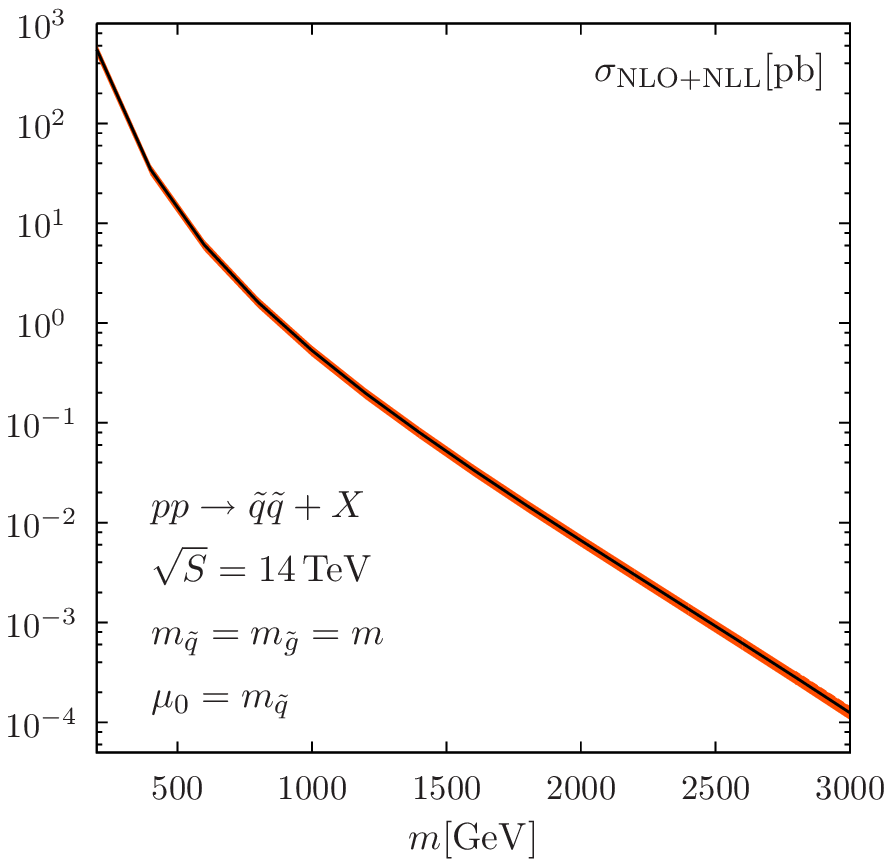,
 width=0.45\columnwidth,angle=0}
\epsfig{file=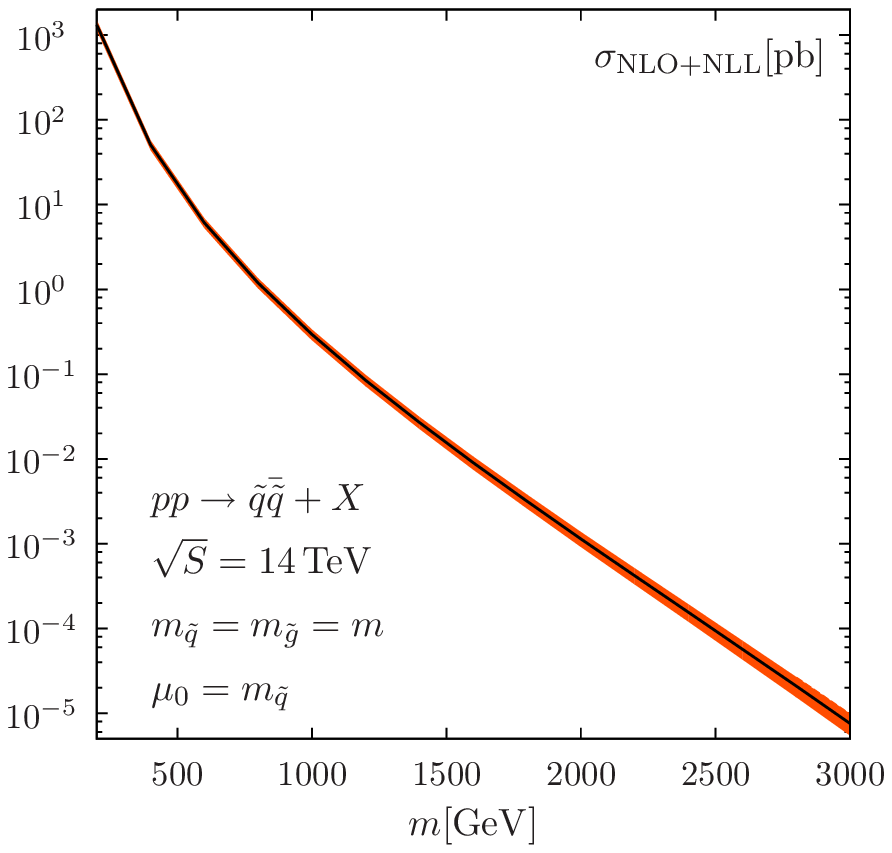,
 width=0.45\columnwidth,angle=0}\\[5mm]
\epsfig{file=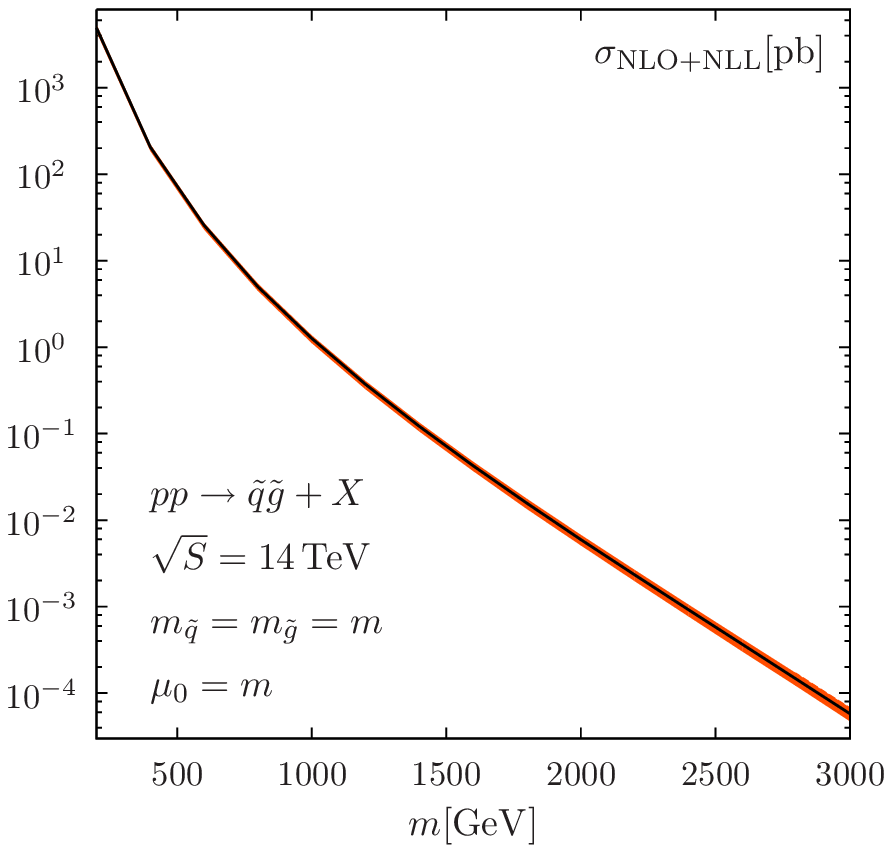,
 width=0.45\columnwidth,angle=0}
\epsfig{file=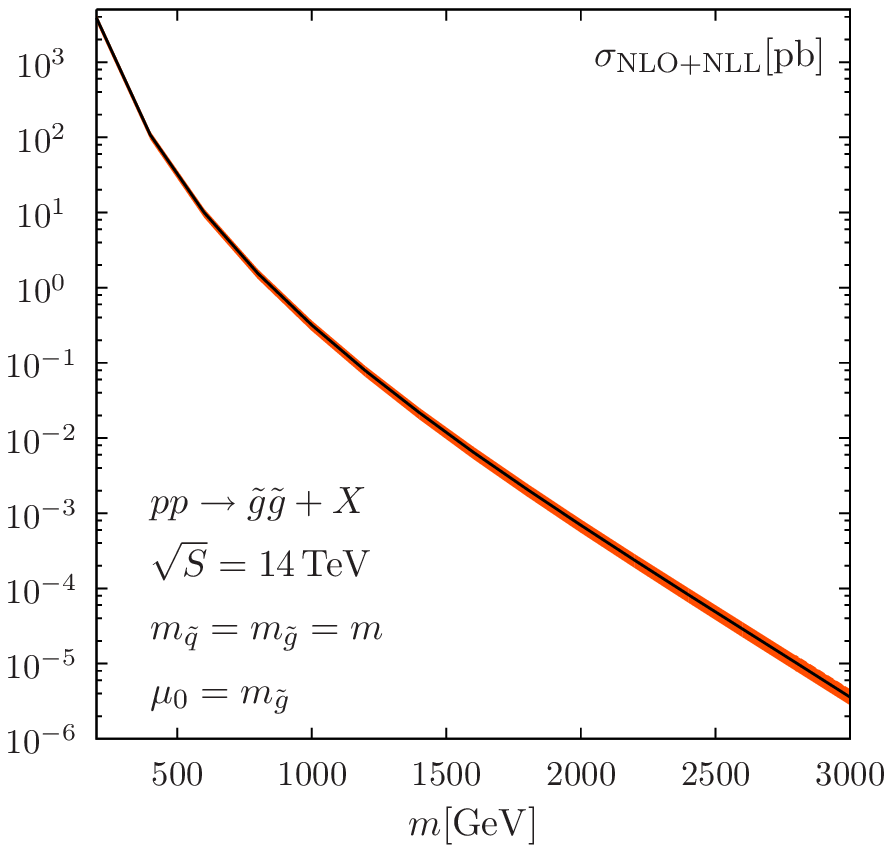,
 width=0.45\columnwidth,angle=0}\\[5mm]
\epsfig{file=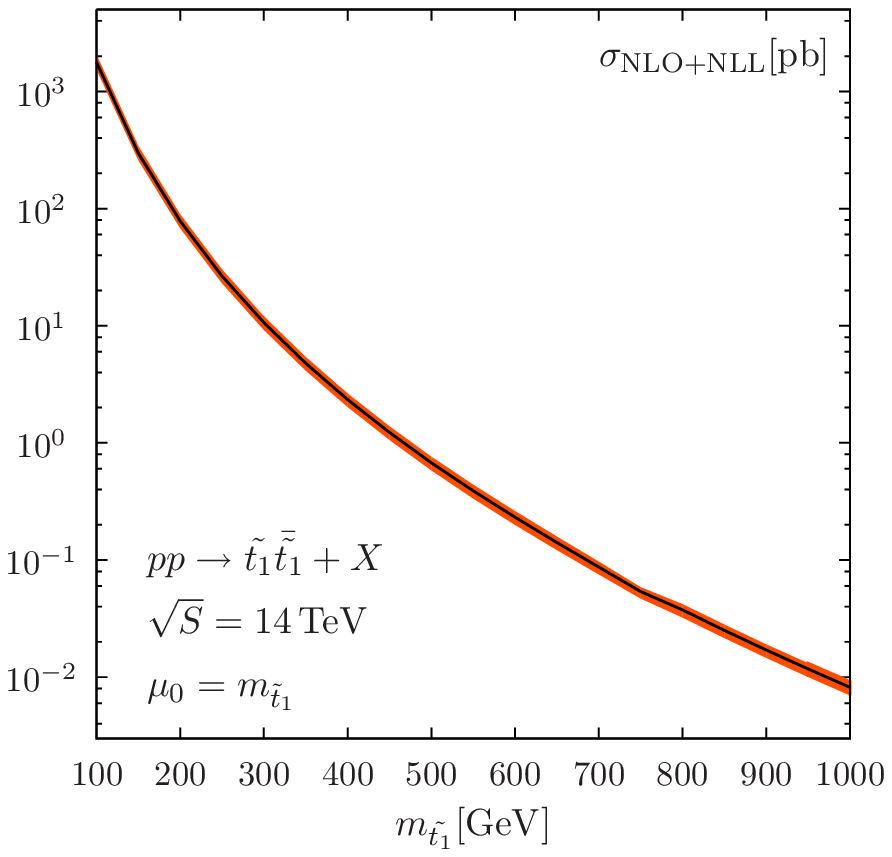, 
 width=0.45\columnwidth, angle=0}
\end{center}
\caption{The NLO+NLL SUSY-QCD cross section for the individual squark
  and gluino pair-production processes at the LHC with 14\,TeV, $pp
  \to \tilde{q}\tilde{q}\,, \tilde{q}\bar{\tilde{q}}\,,
  \tilde{q}\tilde{g}\,, \tilde{g}\tilde{g}+X$ and $pp \to\tilde
  t_1\bar{\tilde t}_1+ X$, as a function of the average sparticle mass
  $m$. The error band includes the 68\% C.L.\ pdf and $\alpha_{\rm s}$
  error, added in quadrature, and the error from scale variation in
  the range $m/2\le \mu \le 2m$ added linearly to the combined pdf and
  $\alpha_{\rm s}$ uncertainty.\label{fig:xs_lhc14}}
\end{figure}

\begin{figure}
\begin{center}
\epsfig{file=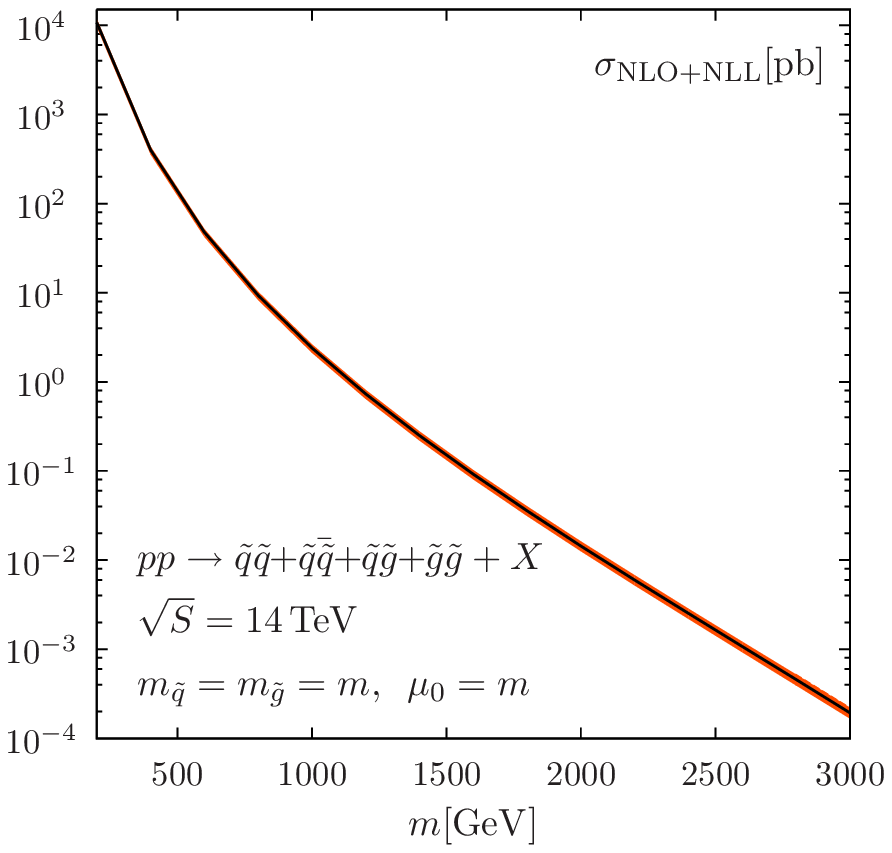,
  width=0.55\columnwidth,angle=0}
\caption{The NLO+NLL SUSY-QCD cross section for inclusive squark and
  gluino pair-production at the LHC with 14\,TeV, $pp \to
  \tilde{q}\tilde{q} + \tilde{q}\bar{\tilde{q}} + \tilde{q}\tilde{g}+
  \tilde{g}\tilde{g}+X$, as a function of the average sparticle mass
  $m$. The error band includes the 68\% C.L.\ pdf and $\alpha_{\rm s}$
  error, added in quadrature, and the error from scale variation in
  the range $m/2\le \mu \le 2m$ added linearly to the combined pdf and
  $\alpha_{\rm s}$ uncertainty.\label{fig:xs_sum_lhc14}}
\end{center}
\end{figure}

\begin{figure}
\begin{center}
\hspace*{-5mm}\epsfig{file=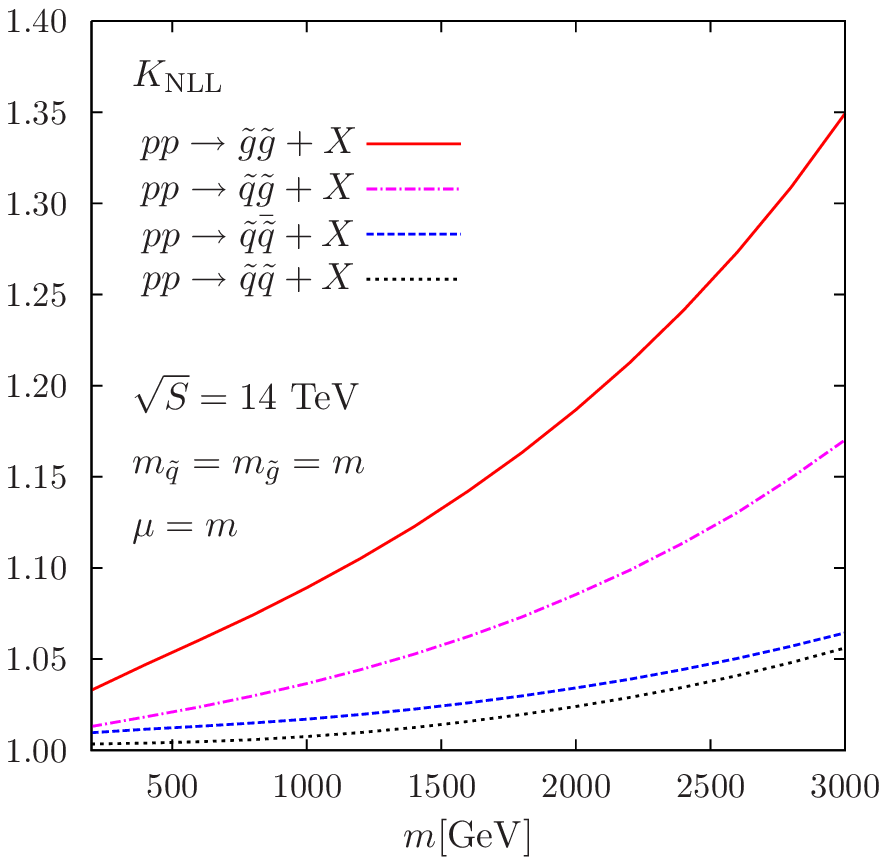, width=0.48\columnwidth,angle=0}
\hspace*{5mm}\epsfig{file=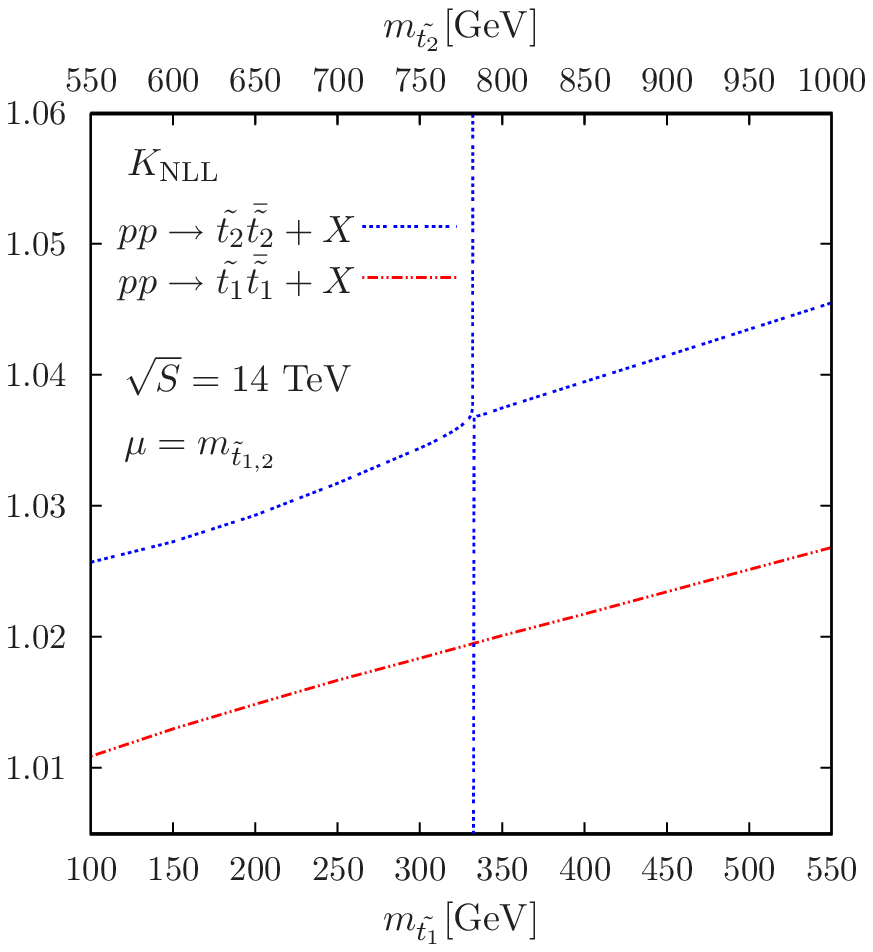, width=0.48\columnwidth,angle=0}
\caption{The NLL $K$-factor $K_{\rm
  NLL}=\sigma_{\rm NLO+NLL}/\sigma_{\rm NLO}$ for the individual squark and gluino
 pair-production processes at the LHC with 14\,TeV as a function of the average sparticle mass $m$.
\label{fig:knll_lhc14}}
 \end{center}
\end{figure}

\begin{figure}
\begin{center}
\epsfig{file=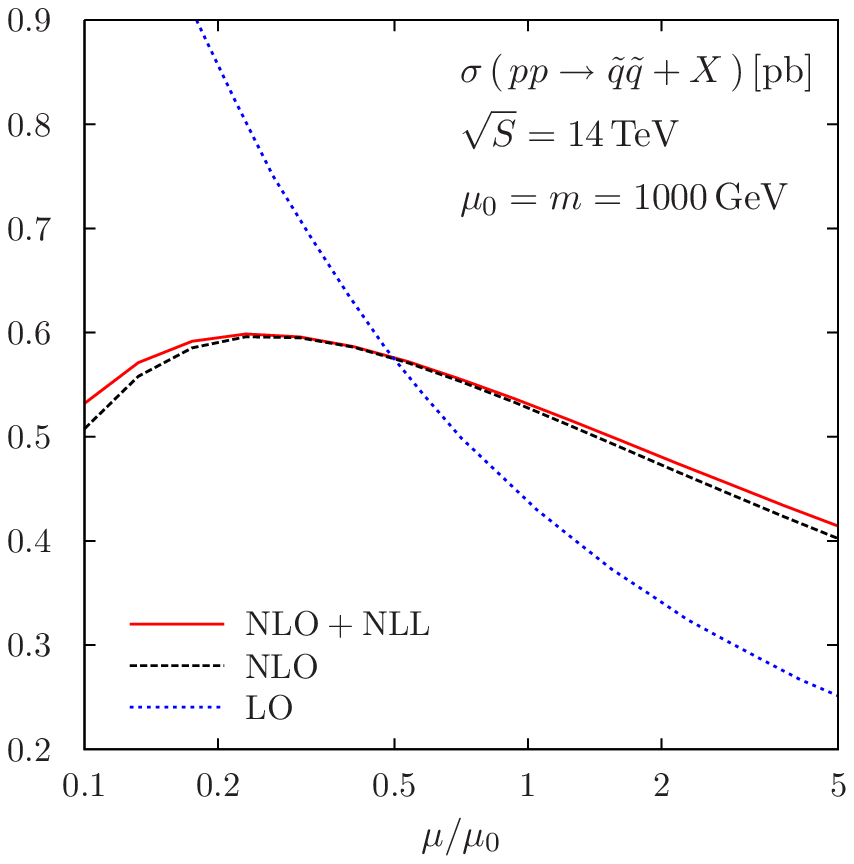, width=0.45\columnwidth,angle=0}
\epsfig{file=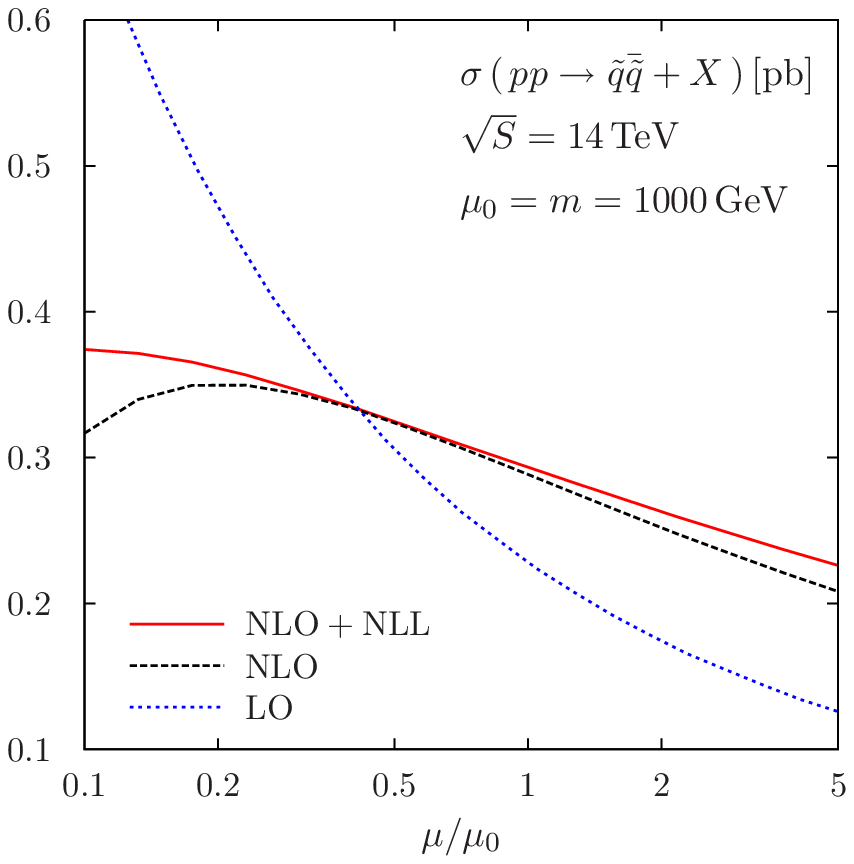, width=0.45\columnwidth,angle=0}\\[5mm]
\epsfig{file=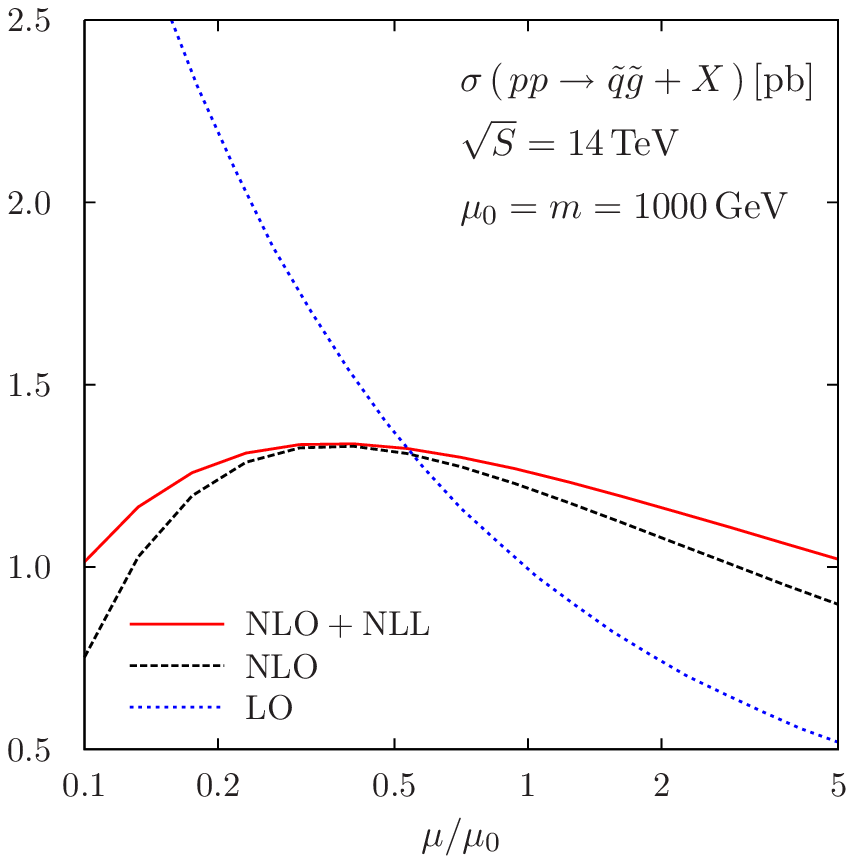, width=0.45\columnwidth, angle=0}
\epsfig{file=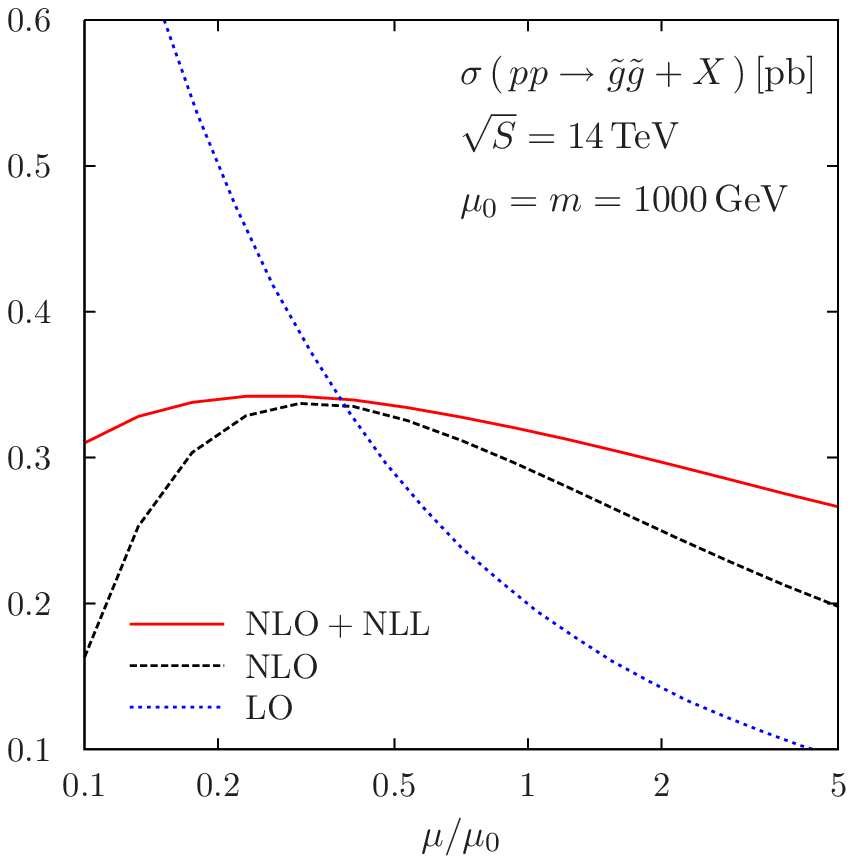, width=0.45\columnwidth, angle=0}\\[5mm]
\epsfig{file=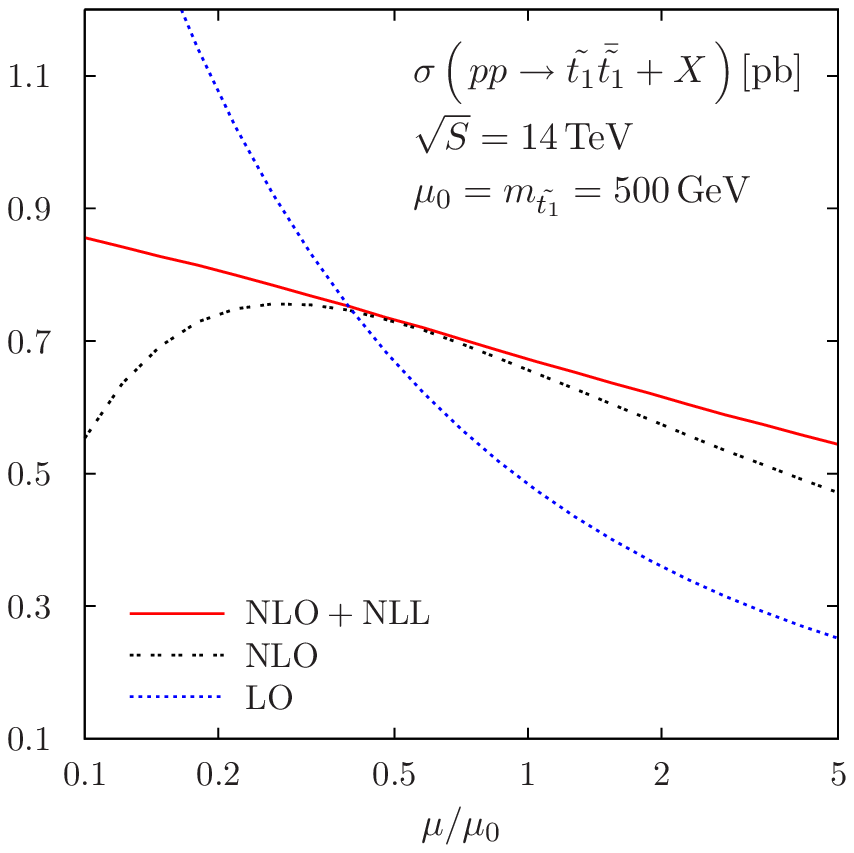, width=0.45\columnwidth, angle=0}
\end{center}
\caption{The scale dependence of the LO, NLO and NLO+NLL cross sections for the individual 
 squark and gluino pair-production processes at the LHC with 14\,TeV.
 The squark and gluino masses have been set equal $m_{\tilde{q}} = m_{\tilde{g}} = m$ in the upper four plots. 
\label{fig:scale_lhc14}}
\end{figure}

SUSY particles with masses in the multi-TeV region can be probed at
the LHC running at or near its design energy of
14\,TeV.\cite{Aad:2009wy,Bayatian:2006zz} To complete this review and
to show how the impact of the higher-order corrections depends on the
collider energy, we also present predictions for squark and gluino
production at the LHC with 14\,TeV. Our cross section predictions at
NLO+NLL including the full theory uncertainty, consisting of the 68\%
C.L.\ pdf and $\alpha_{\rm s}$ error, added in quadrature, combined
linearly with the scale variation error, are shown in
Figs.\,\ref{fig:xs_lhc14} and \ref{fig:xs_sum_lhc14}. As for the LHC
at 7 TeV, the impact of NLL threshold resummation is particularly
significant for gluino-pair and squark-gluino production, with NLL
corrections of about 30\% for $\tilde{g}\tilde{g}$ final states and
squark and gluino masses of 2.5\,TeV, see
Fig.\,\ref{fig:knll_lhc14}. Finally, in Fig.\,\ref{fig:scale_lhc14} we
show the LO, NLO and NLO+NLL scale dependence for squark and gluino
masses of 1\,TeV and $m_{\tilde{t}_1} = 500$\,GeV. The results are
qualitatively very similar to those obtained for the LHC with 7\,TeV,
i.e.\ we find a strong reduction of the scale dependence when
including the high-order corrections, and a very good convergence of
the perturbative series at scales near $\mu = m/2$.

\section{Conclusions}
\label{sec:conclusions}
Precise theoretical predictions for sparticle cross sections are
essential for the interpretation of current and future searches for
supersymmetry at hadron colliders. The inclusion of higher-order
SUSY-QCD corrections reduces the scale uncertainty substantially. The
higher-order terms also increase the size of the cross section with
respect to the LO prediction for renormalization and factorization
scales near the average mass of the produced SUSY particles. Thus, the
SUSY-QCD corrections have a significant impact on the extraction of
SUSY mass bounds from experimental cross section limits, and would
lead to a much more accurate determination of SUSY parameters like
masses and couplings in the case of discovery.

In this review we have presented the state-of-the-art SUSY-QCD
predictions for squark and gluino hadroproduction cross sections at
the Tevatron and the LHC at 7 and 14\,TeV centre-of-mass energy,
including NLO corrections and NLL threshold resummation. We have
discussed the impact of the SUSY-QCD corrections on the cross sections
and have provided an estimate of the theoretical uncertainty due to
scale variation, parton distribution functions, and the strong
coupling $\alpha_{\rm s}$. Special emphasis has been placed on the
predictions for the LHC at 7\,TeV energy, which are of immediate
relevance for ongoing SUSY searches.

The effect of soft-gluon resummation is most pronounced for processes
with initial-state gluons and final-state gluinos, which involve a
large colour charge. Specifically, at the LHC with 7\,TeV we find an
increase of the cross-section prediction of up to 20\% for sparticle
masses around 1\,TeV when going from NLO to NLO+NLL, depending in
detail on the final state and the ratio of squark to gluino
masses. Furthermore, the scale uncertainty is reduced significantly at
NLO+NLL accuracy over the full range of sparticle masses relevant for
hadron collider searches, with a remaining scale uncertainty of less
than 10\%. We have furthermore presented estimates for the overall
theory uncertainty including the 68\% C.L.\ pdf and $\alpha_{\rm s}$
error added in quadrature, combined linearly with the scale variation
error. Even though the pdf and $\alpha_{\rm s}$ uncertainty is
significant, the inclusion of threshold resummation leads to a
sizeable reduction of the overall theory uncertainty, which is in
general 20\% or smaller for NLO+NLL cross sections calculated with
2008 NLO MSTW pdfs.

The NLO+NLL cross sections presented in this paper constitute the
state-of-the-art QCD predictions for squark and gluino production in
the MSSM, and provide therefore, we think, the optimum theoretical
basis to interpret current and future searches for supersymmetry at
the Tevatron and the LHC.

\section*{Acknowledgments}
\noindent This work has been supported in part by the Helmholtz
Alliance ``Physics at the Terascale'', the DFG Graduiertenkolleg
``Elementary Particle Physics at the TeV Scale'', the Foundation for
Fundamental Research of Matter (FOM) program "Theoretical Particle
Physics in the Era of the LHC", the National Organization for
Scientific Research (NWO), the DFG SFB/TR9 ``Computational Particle
Physics'', the European Community's Marie-Curie Research Training
Network under contract MRTN-CT-2006-035505 ``Tools and Precision
Calculations for Physics Discoveries at Colliders'' and by the Polish
Ministry of Science and Higher Education, grant MNiSW no. N202 249235.

\clearpage

\newpage


\clearpage

\begin{sidewaystable}
\vspace*{120mm}
\tbl{The LO, NLO and NLO+NLL cross sections for squark-antisquark and
  squark-squark production at the LHC with 7\,TeV, including errors
  due to scale variation ($\Delta\sigma_{\mu}$) in the range
  $m_{\tilde{q}}/2 \le \mu \le 2m_{\tilde{q}}$.  Results are shown for
  the mass ratio $r=m_{\tilde{g}}/m_{\tilde{q}}=1$ and for two pdf
  parametrizations (MSTW08 and CT10) with the corresponding 68\% C.L.\
  pdf error estimates ($\Delta{\rm{pdf}}$) and $\alpha_{\rm
    s}$-uncertainties ($\Delta{\alpha_{\rm s}}$). Note that
  the $\Delta{\rm{pdf}}$ and $\Delta{\alpha_{\rm s}}$ uncertainties are given as
  relative errors, as opposed to the absolute values of the scale
  variation errors. }
{\begin{tabular}{c|c|c|c||c|c|c}
 \multicolumn{7}{c}{\normalsize $p{p} \to \tilde{q}\bar{\tilde{q}}$ at $\sqrt{S}=7\TeV$}\\[0.5mm]\hline 
 \multicolumn{1}{c}{} & \multicolumn{3}{c||}{MSTW2008} & \multicolumn{3}{c}{CT10} \\[0.5mm] \hline
$m \; [\mathrm{GeV}]$ & 200 & 700 & 1200 & 200 & 700 & 1200   \\  \hline \hline
$\left(\sigma\pm\Delta\sigma_{\mu}\right)_\mathrm{LO} \; [\mathrm{pb}]$ 
& $195^{+68}_{-47} $
& $0.116^{+0.046}_{-0.030} $
& $\left(7.63^{+3.31}_{-2.15}\right)\times 10^{-4}$
& $165^{+54}_{-37}$
& $\left(9.63^{+3.48}_{-2.39}\right)\times10^{-2}$
& $\left(6.09^{+2.44}_{-1.64}\right)\times10^{-4}$\\[0.5mm]  \hline 
$\left(\sigma\pm\Delta\sigma_{\mu}\right)_\mathrm{NLO} \; [\mathrm{pb}]$ 
& $264^{+36}_{-35}$
& $0.143^{+0.022}_{-0.021}$
& $\left(8.31^{+1.59}_{-1.45}\right)\times10^{-4}$
& $250^{+33}_{-32}$
& $0.141^{+0.021}_{-0.021}$
& $\left(9.65^{+1.80}_{-1.66}\right)\times10^{-4}$\\[0.5mm]  \hline
$\left(\sigma\pm\Delta\sigma_{\mu}\right)_\mathrm{NLO+NLL} \; [\mathrm{pb}]$ 
& $ 267^{+33}_{-30}$
& $ 0.146^{+0.019}_{-0.018}$
& $ \left(8.73^{+1.28}_{-1.16}\right)\times10^{-4}$
& $ 253^{+30}_{-27}$
& $ 0.144^{+0.018}_{-0.017}$
& $\left(10.11^{+1.46}_{-1.32}\right)\times10^{-4}$
\\[0.5mm]  \hline
$\Delta\mathrm{pdf}_\mathrm{NLO} \; [\%]$ 
& $^{+1.5}_ {-1.6}$ 
& $^{+5.2}_ {-4.7}$ 
& $^{+11}_{-11}$ 
& $^{+2.9}_ {-2.4}$ 
& $^{+12}_ {-8}$ 
& $^{+37}_{-16}$\\[0.5mm]  \hline 
$\Delta{\alpha_{\rm s}}_\mathrm{NLO} \; [\%]$ 
& $^{+2.4}_{-2.9}$ 
& $^{+2.2}_{-2.4}$ 
& $^{+3.6}_{-3.2}$ 
& $^{+2.4}_{-2.3}$ 
& $^{+3.1}_{-2.6}$ 
& $^{+5.9}_{-4.2}$\\[0.5mm]  \hline 
$\mathrm{K}_{\mathrm{NLO}}$ 
& 1.35 	 	
& 1.23
& 1.09
& 1.51
& 1.46
& 1.59
\\[0.5mm] \hline 
$\mathrm{K}_{\mathrm{NLL}}$ 
& 1.01 
& 1.03 
& 1.05  
& 1.01 
& 1.02
& 1.05\\
 \multicolumn{7}{c}{}\\[0.5mm]
 \multicolumn{7}{c}{\normalsize $p{p} \to \tilde{q}\tilde{{q}}$ at $\sqrt{S}=7\TeV$}\\[0.5mm]\hline 
 \multicolumn{1}{c}{} & \multicolumn{3}{c||}{MSTW2008} & \multicolumn{3}{c}{CT10} \\[0.5mm] \hline
$m \; [\mathrm{GeV}]$ & 200 & 700 & 1200 & 200 & 700 & 1200   \\  \hline \hline
$\left(\sigma\pm\Delta\sigma_{\mu}\right)_\mathrm{LO} \; [\mathrm{pb}]$ 
&$142^{+43}_{-30}$
&$0.349^{+0.127}_{-0.087}$
&$\left(6.06^{+2.48}_{-1.65}\right)\times10^{-3}$
&$128^{+35}_{-26}$
&$0.354^{+0.120}_{-0.084}$
&$\left(6.62^{+2.54}_{-1.72}\right)\times10^{-3}$
\\[0.5mm]  \hline
$\left(\sigma\pm\Delta\sigma_{\mu}\right)_\mathrm{NLO} \; [\mathrm{pb}]$ 
&$175^{+18}_{-19}$
&$0.417^{+0.046}_{-0.051}$
&$\left(7.22^{+1.02}_{-1.06}\right)\times10^{-3}$
&$170^{+17}_{-18}$
&$0.427^{+0.046}_{-0.052}$
&$\left(7.81^{+1.08}_{-1.13}\right)\times10^{-3}$
\\[0.5mm]  \hline
$\left(\sigma\pm\Delta\sigma_{\mu}\right)_\mathrm{NLO+NLL} \; [\mathrm{pb}]$ 
&$175^{+18}_{-18}$
&$0.423	^{+0.042}_{-0.047}$
&$\left(7.50^{+0.85}_{-0.91}\right)\times10^{-3}$
&$171^{+17}_{-17}$
&$0.432^{+0.042}_{-0.047}$
&$\left(8.09^{+0.91}_{-0.99}\right)\times10^{-3}$
\\[0.5mm]  \hline
$\Delta\mathrm{pdf}_\mathrm{NLO} \; [\%]$ 
& $^{+2.1}_ {-1.5}$ 
& $^{+3.4}_ {-2.4}$ 
& $^{+3.7}_ {-2.8}$ 
& $^{+1.4}_ {-1.4}$ 
& $^{+2.4}_ {-2.1}$ 
& $^{+4.3}_ {-3.5}$\\[0.5mm]  \hline 
$\Delta{\alpha_{\rm s}}_\mathrm{NLO} \; [\%]$ 
& $^{+2.3}_{-2.7}$ 
& $^{+0.7}_{-0.9}$ 
& $^{+0.3}_{-0.2}$ 
& $^{+2.2}_{-2.2}$ 
& $^{+0.8}_{-0.8}$ 
& $^{< +0.1}_{< -0.1}$\\[0.5mm]  \hline 
$\mathrm{K}_{\mathrm{NLO}}$ 
& 1.23 	 	
& 1.20
& 1.19
& 1.33
& 1.21
& 1.18
\\[0.5mm] \hline 
$\mathrm{K}_{\mathrm{NLL}}$ 
& 1.00
& 1.01
& 1.04 
& 1.00 
& 1.01
& 1.04
\end{tabular}}
\label{tbl1}
\end{sidewaystable}

\begin{sidewaystable}
 \vspace*{120mm}
\tbl{The LO, NLO and NLO+NLL cross sections for gluino-gluino and
  squark-gluino production at the LHC with 7\,TeV, including errors
  due to scale variation ($\Delta\sigma_{\mu}$) in the range $m/2 \le
  \mu \le 2m$, where $m$ denotes the average sparticle mass. Results
  are shown for the mass ratio $r=m_{\tilde{g}}/m_{\tilde{q}}=1$ and
  for two pdf parametrizations (MSTW08 and CT10) with the
  corresponding 68\% C.L.\ pdf error estimates ($\Delta{\rm{pdf}}$)
  and $\alpha_{\rm s}$-uncertainties ($\Delta{\alpha_{\rm s}}$).  Note that
  the $\Delta{\rm{pdf}}$ and $\Delta{\alpha_{\rm s}}$ uncertainties are given as
  relative errors, as opposed to the absolute values of the scale
  variation errors.}
 {\begin{tabular}{c|c|c|c||c|c|c }
 \multicolumn{7}{c}{\normalsize $p{p} \to \tilde{g}\tilde{{g}}$ at $\sqrt{S}=7\TeV$}\\[0.5mm]\hline 
 \multicolumn{1}{c}{} & \multicolumn{3}{c||}{MSTW2008} & \multicolumn{3}{c}{CT10} \\[0.5mm] \hline
$m \; [\mathrm{GeV}]$ & 200 & 700 & 1200 & 200 & 700 & 1200   \\  \hline \hline
$\hspace{-0.1cm}\left(\sigma\pm\Delta\sigma_{\mu}\right)_\mathrm{LO} \; [\mathrm{pb}]$ 
&$420^{+184}_{-118}\hspace{-0.1cm}$
&$\hspace{-0.1cm}\left(7.51^{+3.98}_{-2.41}\right)\hspace{-0.1cm}\times\hspace{-0.1cm}10^{-2}\hspace{-0.1cm}$
&$\hspace{-0.1cm}\left(2.52^{+1.50}_{-0.87}\right)\hspace{-0.1cm}\times\hspace{-0.1cm}10^{-4}\hspace{-0.1cm}$
&$340^{+139}_{-92}\hspace{-0.1cm}$
&$\hspace{-0.1cm}\left(4.88^{+2.19}_{-1.40}\right)\hspace{-0.1cm}\times\hspace{-0.1cm}10^{-2}\hspace{-0.1cm}$
&$\hspace{-0.1cm}\left(1.48^{+0.69}_{-0.44}\right)\hspace{-0.1cm}\times\hspace{-0.1cm}10^{-4}\hspace{-0.1cm}$\\[0.5mm]  \hline 
$\hspace{-0.1cm}\left(\sigma\pm\Delta\sigma_{\mu}\right)_\mathrm{NLO} \; [\mathrm{pb}]$ 
&$576^{+72}_{-81}\hspace{-0.1cm}$
&$0.114^{+0.019}_{-0.020}\hspace{-0.1cm}$
&$\hspace{-0.1cm}\left(4.12^{+0.91}_{-0.88}\right)\hspace{-0.1cm}\times\hspace{-0.1cm}10^{-4}\hspace{-0.1cm}$
&$541^{+65}_{-74}\hspace{-0.1cm}$
&$0.123^{+0.019}_{-0.021}\hspace{-0.1cm}$
&$\hspace{-0.1cm}\left(5.72^{+1.25}_{-1.23}\right)\hspace{-0.1cm}\times\hspace{-0.1cm}10^{-4}\hspace{-0.1cm}$
\\[0.5mm]  \hline
$\hspace{-0.1cm}\left(\sigma\pm\Delta\sigma_{\mu}\right)_\mathrm{NLO+NLL} \; [\mathrm{pb}]\hspace{-0.1cm}$ 
&$606^{+49}_{-52}\hspace{-0.1cm}$
&$0.130^{+0.008}_{-0.010}\hspace{-0.1cm}$
&$\hspace{-0.1cm}\left(5.25^{+0.26}_{-0.34}\right)\hspace{-0.1cm}\times\hspace{-0.1cm}10^{-4}\hspace{-0.1cm}$
&$569^{+44}_{-46}\hspace{-0.1cm}$
&$0.138^{+0.008}_{-0.011}\hspace{-0.1cm}$
&$\hspace{-0.1cm}\left(7.08^{+0.43}_{-0.55}\right)\hspace{-0.1cm}\times\hspace{-0.1cm}10^{-4}\hspace{-0.1cm}$
\\[0.5mm]  \hline
$\Delta\mathrm{pdf}_\mathrm{NLO} \; [\%]$ 
& $^{+3.8}_ {-4.7}$ 
& $^{+12}_  {-13}$ 
& $^{+22}_  {-21}$ 
& $^{+6.4}_ {-5.4}$ 
& $^{+30}_  {-18}$ 
& $^{+87}_  {-35}$\\[0.5mm]  \hline 
$\Delta{\alpha_{\rm s}}_\mathrm{NLO} \; [\%]$ 
& $^{+2.3}_{-2.8}$ 
& $^{+4.4}_{-4.6}$ 
& $^{+7.6}_{-6.9}$ 
& $^{+2.6}_{-2.4}$ 
& $^{+6.1}_{-4.9}$ 
& $^{+13}_ {-9}$\\[0.5mm]  \hline 
$\mathrm{K}_{\mathrm{NLO}}$ 
& 1.37 	 	
& 1.52
& 1.64
& 1.59
& 2.52
& 3.86
\\[0.5mm] \hline 
$\mathrm{K}_{\mathrm{NLL}}$ 
& 1.05 
& 1.14 
& 1.27  
& 1.05 
& 1.13
& 1.24\\
 \multicolumn{7}{c}{}\\[0.5mm]
 \multicolumn{7}{c}{\normalsize $p{p} \to \tilde{q}{\tilde{g}}$ at $\sqrt{S}=7\TeV$}\\[0.5mm]\hline 
 \multicolumn{1}{c}{} & \multicolumn{3}{c||}{MSTW2008} & \multicolumn{3}{c}{CT10} \\[0.5mm] \hline
$m\; [\mathrm{GeV}]$ & 200 & 700 & 1200 & 200 & 700 & 1200   \\  \hline \hline
$\hspace{-0.1cm}\left(\sigma\pm\Delta\sigma_{\mu}\right)_\mathrm{LO} \; [\mathrm{pb}]$
&$917^{+330}_{-225}\hspace{-0.1cm}$
&$0.515^{+0.228}_{-0.147}\hspace{-0.1cm}$
&$\hspace{-0.1cm}\left(3.66^{+1.84}_{-1.14}\right)\hspace{-0.1cm}\times\hspace{-0.1cm} 10^{-3}\hspace{-0.1cm}$
&$824^{+280}_{-194}\hspace{-0.1cm}$
&$0.434^{+0.174}_{-0.116}\hspace{-0.1cm}$
&$\hspace{-0.1cm}\left(2.93^{+1.27}_{-0.83}\right)\hspace{-0.1cm}\times\hspace{-0.1cm} 10^{-3}\hspace{-0.1cm}$
\\[0.5mm]  \hline 
$\hspace{-0.1cm}\left(\sigma\pm\Delta\sigma_{\mu}\right)_\mathrm{NLO} \; [\mathrm{pb}]$ 
&$\hspace{-0.1cm}\left(1.07^{+0.09}_{-0.12}\right)\hspace{-0.1cm}\times\hspace{-0.1cm} 10^{3}\hspace{-0.1cm}$
&$0.642^{+0.069}_{-0.088}\hspace{-0.1cm}$
&$\hspace{-0.1cm}\left(4.81^{+0.73}_{-0.81}\right)\hspace{-0.1cm}\times\hspace{-0.1cm} 10^{-3}\hspace{-0.1cm}$
&$\hspace{-0.1cm}\left(1.02^{+0.08}_{-0.11}\right)\hspace{-0.1cm}\times\hspace{-0.1cm} 10^{3}\hspace{-0.1cm}$
&$0.657^{+0.068}_{-0.088}\hspace{-0.1cm}$
&$\hspace{-0.1cm}\left(5.74^{+0.84}_{-0.96}\right)\hspace{-0.1cm}\times\hspace{-0.1cm} 10^{-3}\hspace{-0.1cm}$
\\[0.5mm]  \hline
$\hspace{-0.1cm}\left(\sigma\pm\Delta\sigma_{\mu}\right)_\mathrm{NLO+NLL} \; [\mathrm{pb}]$ 
&$\hspace{-0.1cm}\left(1.09^{+0.07}_{-0.09}\right)\hspace{-0.1cm}\times\hspace{-0.1cm}10^{3}\hspace{-0.1cm}$
&$0.680^{+0.042}_{-0.061}\hspace{-0.1cm}$
&$\hspace{-0.1cm}\left(5.43^{+0.35}_{-0.49}\right)\hspace{-0.1cm}\times\hspace{-0.1cm}10^{-3}\hspace{-0.1cm}$
&$\hspace{-0.1cm}\left(1.04^{+0.07}_{-0.09}\right)\hspace{-0.1cm}\times\hspace{-0.1cm} 10^{3}\hspace{-0.1cm}$
&$0.693^{+0.042}_{-0.061}\hspace{-0.1cm}$
&$\hspace{-0.1cm}\left(6.39^{+0.43}_{-0.61}\right)\hspace{-0.1cm}\times\hspace{-0.1cm}10^{-3}\hspace{-0.1cm}$
\\[0.5mm]  \hline
$\Delta\mathrm{pdf}_\mathrm{NLO} \; [\%]$ 
& $^{+0.9}_{-1.0}$ 
& $^{+4.7}_ {-4.9}$ 
& $^{+9.7}_ {-9.3}$ 
& $^{+1.8}_ {-1.5}$ 
& $^{+11}_  {-8}$ 
& $^{+32}_  {-16 }$\\[0.5mm]  \hline 
$\Delta{\alpha_{\rm s}}_\mathrm{NLO} \; [\%]$ 
& $^{+1.8}_{-2.3}$ 
& $^{+2.0}_{-2.3}$ 
& $^{+3.1}_{-3.0}$ 
& $^{+1.9}_{-1.8}$ 
& $^{+2.6}_{-2.2}$ 
& $^{+5.1}_{-4.1}$\\[0.5mm]  \hline 
$\mathrm{K}_{\mathrm{NLO}}$ 
& 1.16 	 	
& 1.25
& 1.32
& 1.24
& 1.52
& 1.96
\\[0.5mm] \hline 
$\mathrm{K}_{\mathrm{NLL}}$ 
& 1.02 
& 1.06 
& 1.13  
& 1.02 
& 1.06
& 1.11
\end{tabular}}
\label{tbl2}
\end{sidewaystable}

\begin{sidewaystable}
 \vspace*{120mm}
\tbl{The LO, NLO and NLO+NLL cross sections for stop-antistop
  production at the LHC with 7\,TeV, including errors due to scale
  variation ($\Delta\sigma_{\mu}$) in the range $m_{\tilde{t}}/2 \le
  \mu \le 2m_{\tilde{t}}$. The SUSY parameters $m_{\tilde{g}}$,
  $m_{\tilde{q}}$ and $\theta_{\tilde{t}}$ have been set to the SPS1a'
  benchmark values. Results are shown for two pdf parametrizations
  (MSTW08 and CT10) with the corresponding 68\% C.L.\ pdf error
  estimates ($\Delta{\rm{pdf}}$) and $\alpha_{\rm s}$-uncertainties
  ($\Delta{\alpha_{\rm s}}$). Note that
  the $\Delta{\rm{pdf}}$ and $\Delta{\alpha_{\rm s}}$ uncertainties are given as
  relative errors, as opposed to the absolute values of the scale
  variation errors. }
{\begin{tabular}{c|c|c||c|c}
 \multicolumn{5}{c}{\normalsize $p{p} \to \tilde{t_1}\bar{\tilde{t_1}}$ at $\sqrt{S}=7\TeV$}
 \\[0.5mm]  \hline 
 \multicolumn{1}{c}{} & \multicolumn{2}{c||}{MSTW2008} & \multicolumn{2}{c}{CT10} \\[0.5mm] \hline
$m_{\tilde{t_1}} \; [\mathrm{GeV}]$ & 100 & 400 & 100 & 400  \\[0.5mm]  \hline \hline
$(\sigma\pm\Delta\sigma_{\mu})_\mathrm{LO} \; [\mathrm{pb}]$ 
& $305 ^{+114}_{-77}$ 
& $0.156 ^{+0.070}_{-0.044}$ 
& $265 ^{+95}_{-65}$ 
& $0.119 ^{+0.048}_{-0.032}$\\[0.5mm]  \hline 
$(\sigma\pm\Delta\sigma_{\mu})_\mathrm{NLO} \; [\mathrm{pb}]$ 
& $416^{+64}_{-59}$ 
& $0.209 ^{+0.027}_{-0.031}$ 
& $389 ^{+58}_{-53}$ 
& $0.202 ^{+0.025}_{-0.029}$\\[0.5mm]  \hline 
$(\sigma\pm\Delta\sigma_{\mu})_\mathrm{NLO+NLL} \; [\mathrm{pb}]$ 
& $423^{+60}_{-46}$ 
& $0.218 ^{+0.020}_{-0.020}$ 
& $395 ^{+54}_{-42}$ 
& $0.209 ^{+0.018}_{-0.019}$\\[0.5mm]  \hline
$\Delta\mathrm{pdf}_\mathrm{NLO} \; [\%]$ 
& $^{+2.0}_ {-2.4}$ 
& $^{+5.8}_  {-6.3}$ 
& $^{+2.7}_  {-2.5}$ 
& $^{+11}_  {-9}$\\[0.5mm]  \hline
$\Delta{\alpha_{\rm s}}_\mathrm{NLO} \; [\%]$ 
& $^{+2.1}_ {-2.5}$ 
& $^{+2.7}_ {-3.1}$ 
& $^{+2.1}_ {-2.1}$ 
& $^{+3.2}_ {-2.8}$\\[0.5mm]  \hline
$\mathrm{K}_{\mathrm{NLO}}$ 
& 1.37 
& 1.34 
& 1.47
& 1.70  \\[0.5mm]  \hline 
$\mathrm{K}_{\mathrm{NLL}}$ 
& 1.02 
& 1.04 
& 1.02 
& 1.04 \\[0.5mm]
 \multicolumn{5}{c}{}\\[0.5mm]
 \multicolumn{5}{c}{\normalsize $p{p} \to \tilde{t_2}\bar{\tilde{t_2}}$ at $\sqrt{S}=7\TeV$}
 \\[0.5mm]  \hline 
 \multicolumn{1}{c}{} & \multicolumn{2}{c||}{MSTW2008} & \multicolumn{2}{c}{CT10} \\[0.5mm] \hline
$m_{\tilde{t_2}} \; [\mathrm{GeV}]$ & 600 & 1000 & 600 & 1000  \\[0.5mm]  \hline \hline
$(\sigma\pm\Delta\sigma_{\mu})_\mathrm{LO} \; [\mathrm{pb}]$ 
& $(9.06 ^{+4.22}_{-2.66})\times 10^{-3}$ 
& $(9.64 ^{+4.83}_{-2.97})\times 10^{-5}$ 
& $(6.63 ^{+2.70}_{-1.78})\times 10^{-3}$ 
& $(6.76 ^{+2.86}_{-1.88})\times 10^{-5}$\\[0.5mm]  \hline 
$(\sigma\pm\Delta\sigma_{\mu})_\mathrm{NLO} \; [\mathrm{pb}]$ 
& $(1.23 ^{+0.18}_{-0.20})\times 10^{-2}$ 
& $(1.17 ^{+0.15}_{-0.19})\times 10^{-4}$ 
& $(1.24 ^{+0.18}_{-0.20})\times 10^{-2}$ 
& $(1.33 ^{+0.17}_{-0.22})\times 10^{-4}$\\[0.5mm]  \hline 
$(\sigma\pm\Delta\sigma_{\mu})_\mathrm{NLO+NLL} \; [\mathrm{pb}]$ 
& $(1.30 ^{+0.13}_{-0.12})\times 10^{-2}$ 
& $(1.31 ^{+0.05}_{-0.09})\times 10^{-4}$ 
& $(1.30 ^{+0.12}_{-0.13})\times 10^{-2}$ 
& $(1.47 ^{+0.06}_{-0.11})\times 10^{-4}$\\[0.5mm]  \hline
$\Delta\mathrm{pdf}_\mathrm{NLO} \; [\%]$ 
& $^{+8.3}_ {-8.7}$ 
& $^{+14}_  {-13}$ 
& $^{+18}_  {-12}$ 
& $^{+42}_  {-20}$\\[0.5mm]  \hline
$\Delta{\alpha_{\rm s}}_\mathrm{NLO} \; [\%]$ 
& $^{+3.2}_ {-3.5}$ 
& $^{+4.3}_  {-4.0}$ 
& $^{+4.2}_  {-3.5}$ 
& $^{+7.0}_  {-5.4}$\\[0.5mm]  \hline
$\mathrm{K}_{\mathrm{NLO}}$ 
& 1.36 
& 1.22 
& 1.87
& 1.97  \\[0.5mm] \hline 
$\mathrm{K}_{\mathrm{NLL}}$ 
& 1.06 
& 1.11 
& 1.05 
& 1.10
\end{tabular}}
\label{tbl3}
\end{sidewaystable}

\end{document}